%% file: main-camera-ready.tex
\documentclass[lettersize,journal]{IEEEtran}
\usepackage{amsmath,amsfonts}
\usepackage{algorithmic}
\usepackage{algorithm}
\usepackage{array}
\usepackage[caption=false,font=normalsize,labelfont=sf,textfont=sf]{subfig}
\usepackage{textcomp}
\usepackage{stfloats}
\usepackage{url}
\usepackage{verbatim}
\usepackage{color,soul}
\usepackage{graphicx}
\usepackage{hyperref}
\usepackage{cite}
\usepackage[nolist]{acronym}
\usepackage{mathtools}
\acrodefplural{HPA}[HPAs]{high power amplifiers}

\input{acronyms.tex}

\DeclareMathOperator{\SNR}{SNR}
\DeclareMathOperator{\SINR}{SINR}
\DeclareMathOperator{\MUI}{MUI}
\DeclareMathOperator{\PAPR}{PAPR}
\DeclareMathOperator{\comm}{comm}
\DeclareMathOperator{\ve}{vec}
\DeclareMathOperator{\Real}{Re}
\DeclareMathOperator{\Imag}{Im}
\DeclareMathOperator{\dB}{dB}

\newcommand\myeqa{\mathrel{\overset{\makebox[0pt]{\mbox{\normalfont\tiny\sffamily (a)}}}{=}}}
\newcommand\myeqb{\mathrel{\overset{\makebox[0pt]{\mbox{\normalfont\tiny\sffamily (b)}}}{=}}}

\newcommand\myeqd{\mathrel{\overset{\makebox[0pt]{\mbox{\normalfont\tiny\sffamily (d)}}}{=}}}

\newcommand\myleqc{\mathrel{\overset{\makebox[0pt]{\mbox{\normalfont\tiny\sffamily (c)}}}{\leq}}}

\usepackage{xcolor,cite,etoolbox}
\makeatletter 
\pretocmd\@bibitem{\color{black}\csname keycolor#1\endcsname}{}{\fail}
\newcommand\citecolor[1]{\@namedef{keycolor#1}{\color{black}}}
\makeatother

\citecolor{chafii2022ten}
\citecolor{marzetta2015massive}
\citecolor{ali2016modeling}
\citecolor{sigmund2022iterative}
\citecolor{ali2018downlink}
\citecolor{liu2021dual}
\citecolor{zhang2020perceptive}
\citecolor{liu2020joint}
\citecolor{kumari2019adaptive}
\citecolor{liu2018toward}
\citecolor{qian2018joint}
\citecolor{xu2020multi}
\citecolor{su2020secure}
\citecolor{ren2022robust}
\citecolor{su2022secure}
\citecolor{bazzi2022outage}
\citecolor{ma2020automotive}
\citecolor{hassanien2019dual}
\citecolor{liu2022integrated}
\citecolor{ma2019wifi}
\citecolor{meilhac2022digital}
\citecolor{mancuso2011reducing}
\citecolor{jiang2008derivation}
\citecolor{wang2021model}
\citecolor{chafii2016necessary}
\citecolor{chafii2014papr}
\citecolor{tsai2011synthesizing}
\citecolor{zhou2020new}
\citecolor{mabrouk2017precoding}
\citecolor{hu2022low}
\citecolor{bazzi2015efficient}
\citecolor{wang2010optimized}
\citecolor{de2008design}
\citecolor{cui2013mimo}
\citecolor{cui2016space}
\citecolor{barros1996new}
\citecolor{tuy1998convex}
\citecolor{chang2014multi}
\citecolor{chan2019performance}
\citecolor{zhang2021overview}
\citecolor{mishra2019toward}
\citecolor{mohammed2013per}
\citecolor{liu2020range}
\citecolor{zhao2020mimo}
\citecolor{boyd2011distributed}
\citecolor{aldayel2016successive}
\citecolor{huang2011robust}
\citecolor{richards2010principles}
\citecolor{smeenk2022localization}
\citecolor{berggren2022joint}
\begin{document}

\title{On Integrated Sensing and Communication Waveforms with Tunable PAPR}

\author{Ahmad Bazzi and Marwa Chafii 
\thanks{Ahmad Bazzi is with the Engineering Division, New York University (NYU) Abu Dhabi, 129188, UAE
(email: \href{ahmad.bazzi@nyu.edu}{ahmad.bazzi@nyu.edu}).

Marwa Chafii is with Engineering Division, New York University (NYU) Abu Dhabi, 129188, UAE and NYU WIRELESS,
NYU Tandon School of Engineering, Brooklyn, 11201, NY, USA (email: \href{marwa.chafii@nyu.edu}{marwa.chafii@nyu.edu}).}
\thanks{Manuscript received xxx}}

\markboth{\MakeLowercase{to appear in} IEEE Transactions on Wireless Communications, 2023}%
{Shell \MakeLowercase{\textit{et al.}}: A Sample Article Using IEEEtran.cls for IEEE Journals}

\IEEEpubid{}

\maketitle

\begin{abstract}
We present a novel approach to the problem of dual-functional radar and communication (DFRC) waveform design with adjustable peak-to-average power ratio (PAPR), while minimizing the multi-user communication interference and maintaining a similarity constraint towards a radar chirp signal. The approach is applicable to generic radar chirp signals and for different constellation sizes. We formulate the waveform design problem as a non convex optimization problem. As a solution, we adopt the alternating direction method of multipliers (ADMM), hence iterating towards a stable waveform for both radar and communication purposes. \textcolor{black}{Additionally, we prove convergence of the proposed method and analyze its computational complexity. Moreover, we offer an extended version of the method to cope with imperfect channel state information (CSI). Finally, we demonstrate its superior performance through simulations, in comparison to state-of-the-art radar-communication waveform designs.}
\end{abstract}


\section{Introduction}
\label{sec:introduction}
\IEEEPARstart{B}y examining possible emerging services and applications, identifying market needs, and pinpointing disruptive technologies, \textcolor{black}{research} has started to put together a speculative image of 6G \cite{chafii2022ten}. Although the deployment of 5G networks is still ongoing, some key paradigms have been \textcolor{black}{identified} as the network's building blocks based on research and the associated implementation. 
\textcolor{black}{\input{actions01/B5G6G.tex}}


 In particular, by leveraging the same transmit signal via a fully-shared transmitter, \ac{DFRC} systems concurrently perform radar and communication operations. Thanks to such \textcolor{black}{an} approach, full collaboration between radar and communication sub-systems \textcolor{black}{can be accomplished}, while only requiring  smaller-size, lower-cost, and \textcolor{black}{less-complex platforms} \cite{liu2021dual}. For more advantages and applications on \ac{DFRC}, the reader is referred to \cite{zhang2020perceptive}. To meet the conflicting demands of communication and sensing, advanced designs for dual functional waveforms are necessary, in addition to the integration and coordination advantages. Furthermore, \textcolor{black}{\ac{MIMO}} design has been widely adopted in \ac{DFRC} systems, due to its advantage of better exploiting the trade-offs between radar and communications, thanks to the spatial degrees of freedom \cite{liu2020joint}. Lately, a significant amount of research has been oriented towards \ac{DFRC} beamforming \cite{liu2018toward}, symbol-level precoding \cite{liu2021dual}, PHY-layer security \cite{su2022secure}, and robust beamforming \cite{bazzi2022outage}. Indeed, \ac{DFRC} designs find attractive applications, such as in the automotive systems \cite{ma2020automotive}, military and defense \cite{hassanien2019dual}, enhanced localization and tracking \cite{liu2022integrated}, human activity recognition \cite{ma2019wifi}.

Another important and favorable requirement that prevails in both radar and communication systems is low \ac{PAPR} transmissions \cite{meilhac2022digital} for energy-efficient purposes, especially when non-linear \textcolor{black}{\acp{HPA}} are integrated within the transmit chain \cite{mancuso2011reducing}. By definition, the \ac{PAPR} \cite{jiang2008derivation} is a random variable that measures the power variations of signals. \textcolor{black}{In principle, low \ac{PAPR} waveforms are desired, as it enables us to tune the \ac{HPA}'s Q-point as close as possible towards the optimal operating point, with no risk of clipping.} Indeed, \ac{PAPR} reduction methods have been studied in depth. \textcolor{black}{For instance, the work in \cite{wang2021model} proposes \ac{PAPR} reduction methods for \ac{OFDM} and \cite{chafii2016necessary} derives necessary conditions for waveforms exhibiting a better \ac{PAPR} than \ac{OFDM}.} Some methods leverage convex optimization to synthesize favorable sequences with low \ac{PAPR} properties and spectral mask constraints\cite{zhou2020new}, whereas others implement \textcolor{black}{baseband pre-distortion methods} to reduce the \ac{PAPR}, \textcolor{black}{at the price of increased} \ac{EVM} \cite{meilhac2022digital}. Pre-coding based \ac{PAPR} reduction techniques \cite{mabrouk2017precoding} also seem to be promising due to their reduced computational complexity, since they are linear to implement without any prior information.

\subsection{\textcolor{black}{Literature Review}}
\textcolor{black}{\input{actions01/literature.tex}}
\subsection{Contributions and Insights}
\label{subsec:contributions}
This paper considers \ac{DFRC} \textcolor{black}{\ac{BS}} waveform design \textcolor{black}{with tunable \ac{PAPR}, intended for \ac{DL}} communication users, while listening to \textcolor{black}{the} received echo of the transmitted signal. We first formulate a non-convex optimization problem that aims at minimizing an important metric for communications, namely the multi-user interference over all communication users, with a given similarity constraint relative to a radar chirp signal and under a given \ac{PAPR} constraint. We adopt the \ac{ADMM} method as our solution to solve the proposed non-convex optimization problem. To that purpose, we have summarized our contributions below
\begin{itemize}
	\item \textbf{\ac{DFRC} waveform design with \ac{PAPR} control}. We propose a \ac{DFRC} waveform optimization framework, capable of multi-user interference minimization, while guaranteeing a similarity constraint relative to a radar chirp waveform. Even more, the waveform optimization framework allows us to control the transmit PAPR, which is a favorable feature in practical PHY layer designs.  
	\item \textbf{Solution via \ac{ADMM}}. Our proposed \ac{DFRC} waveform design problem is a non-convex optimization problem, in which we adopt an \ac{ADMM}-based iterative method as a solution.
	\item \textcolor{black}{\input{actions01/contribution-complexity.tex}}
	\item \textcolor{black}{\input{actions01/contribution-CSI.tex}}
	\vspace{-0.4cm}
	\item \textbf{Extensive simulation results}. In order to highlight the various benefits of the proposed waveform design and the capability of the \ac{ADMM}-based waveform design solution in both radar sensing and multi-user communications, we present extensive simulation results showing the potential and superiority of the proposed design, when compared to state-of-the-art designs.
\end{itemize}
Furthermore, we unveil some important insights, i.e.
\begin{itemize}
	\item The proposed waveform design works for generic constellations, such as $M$-QAM.  
	\item The proposed \ac{ADMM}-based waveform design can achieve the \ac{AWGN} capacity performance, in terms of average achievable sum rate.
	\item \textcolor{black}{\input{actions01/claim-proof.tex}}
\end{itemize}
\subsection{Organization and Notations}
\label{subsec:organization}
\textcolor{black}{\input{actions01/organization.tex}}

\textbf{Notation}: Upper-case and lower-case boldface letters denote matrices and vectors, respectively. $(.)^T$, $(.)^*$ and $(.)^H$ represent the transpose, the conjugate and the transpose-conjugate operators. The statistical expectation is $\mathbb{E}\lbrace . \rbrace$. For any complex number $z \in \mathbb{C}$, the magnitude is $\vert z \vert$, its real part is $\Real(z)$, and  its imaginary part is $\Imag(z)$. The Frobenius norm of matrix $\pmb{X}$ is $\Vert \pmb{X} \Vert$. The matrix $\pmb{I}_N$ is the identity matrix of size $N \times N$. The zero-vector is $\pmb{0}$. The inverse of a square matrix is $\pmb{X}^{-1}$. Furthermore, the vectorization and unvectorization operators are denoted as $\ve$ and $\ve^{-1}$, respectively. In particular, $\ve$ takes an $N\times M$ matrix $\pmb{X}$ as input and returns an $NM \times 1$ vector, by stacking the columns of $\pmb{X}$. We index the $(i,j)^{th}$ entry of matrix $\pmb{A}$ as $\pmb{A}_{i,j}$. For compactness, we denote the $i^{th}$ row of matrix $\pmb{A}$ as $\pmb{A}_{i}$. The all-ones vector of size $N$ is denoted as $\pmb{1}_N$ and $\pmb{o}_k$ is an all-zeros vector, except for its $k^{th}$ entry, which is set to $1$. The Kronecker product is denoted as $\otimes$. The value at the $m^{th}$ iteration of a quantity, say $x$, involved in an iterative-type algorithm is denoted as $x^{(m)}$. \textcolor{black}{For two vectors $\pmb{x}$ and $\pmb{y}$, $\begin{bmatrix}
		\pmb{x} , \pmb{y}
	\end{bmatrix}^{(m)}$ denotes $\begin{bmatrix}
		\pmb{x}^{(m)} , \pmb{y}^{(m)}
	\end{bmatrix}$ }. \textcolor{black}{We use big $\mathcal{O}$ notation, i.e. $\mathcal{O}(g(x))$ to refer to a function $f(x)$ if there exists a constant $K$ such that $\vert f(x) \vert \leq K g(x)$ for every $x$.} 

\section{System Model}
\label{sec:systemmodel}
\begin{figure}[!t]
\centering
\includegraphics[width=3.5in]{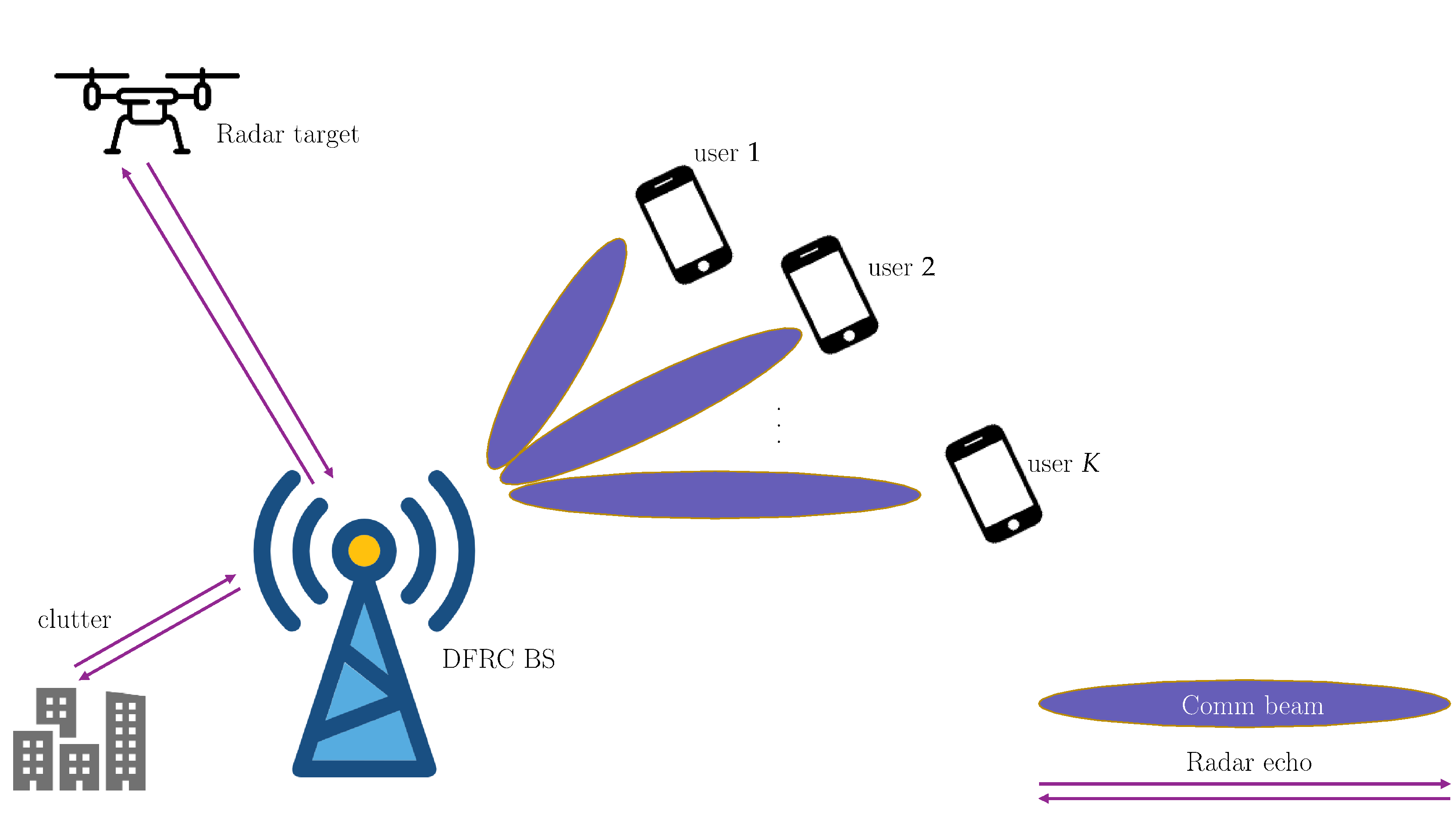}
\caption{
\ac{DFRC} scenario including a \ac{DFRC} \textcolor{black}{\ac{BS}}, an intended target with clutter and $K$ communication users.}
\label{fig:fig_1}
	\end{figure}
\label{sec:system-model}
	Consider a \ac{DFRC} system comprised of a target of interest, $K$ single-antenna communication users, and a \ac{DFRC} \textcolor{black}{\ac{BS}}. The \textcolor{black}{\ac{BS}} is equipped with an antenna array composed of $N$ elements. Fig. \ref{fig:fig_1} depicts a \ac{DFRC} \textcolor{black}{\ac{BS}} broadcasting the same signal vector to communication users and an intended target of interest. Communication users are considered to be located at random positions, whereas the target is supposed to be at a given angle $\theta_0$ from the \ac{DFRC} \textcolor{black}{\ac{BS}}. 
\subsection{Communication System Model}
\label{subsec:comm-sys-model}
A single transmission in the \textcolor{black}{\ac{DL}} sense, \textcolor{black}{initiated} from a \ac{DFRC} \textcolor{black}{\ac{BS}} equipped with $N$ antennas, can be expressed as 
\begin{equation}
	\label{eq:y=Hxpz}
	\pmb{Y}_c = \pmb{H}\pmb{X} + \pmb{Z}_c,
\end{equation}
where $\pmb{Y}_c \in \mathbb{C}^{K \times L}$ is the \textcolor{black}{matrix} of received signals, i.e. the $k^{th}$ row of $\pmb{Y}_c$ is the received sampled waveform at the $k^{th}$ communication user. Furthermore, $L$ denotes the number of time samples of the \ac{DFRC} signal $\pmb{X}$. The channel matrix is given by $\pmb{H} = \begin{bmatrix} \pmb{h}_1 & \pmb{h}_2 & \ldots & \pmb{h}_K \end{bmatrix}^T \in \mathbb{C}^{K \times N}$, and is flat Rayleigh type fading, assumed to be constant during one transmission. Also, the \textcolor{black}{\ac{BS}} assumes full knowledge of the channel $\pmb{H}$. Furthermore, the transmit signal matrix is denoted by $\pmb{X} \in \mathbb{C}^{N \times L}$. Finally, the vector $\pmb{Z}_c \in \mathbb{C}^{K \times L}$ is background noise, where each column is white Gaussian i.i.d with zero mean and a multiple of identity covariance matrix as $\mathcal{N}(0,\sigma_c^2 \pmb{I}_K)$. Note that the signal $\pmb{X}$ is used for both communication and sensing tasks \cite{liu2018toward}. 
\subsection{Radar System Model}
\label{subsec:radar-system-model}
The radar functionality in the \ac{DFRC} system leverages the same transmit signal as the one used for the communication system model, namely $\pmb{X}$. In that way, and using only one time slot, the \ac{DFRC} is capable of achieving dual sensing and communication functionalities. Thanks to the sensing capability of the \ac{DFRC} \textcolor{black}{\ac{BS}}, and assuming a colocated mono-static \ac{MIMO} radar setting \cite{zhang2021overview}
\textcolor{black}{\input{actions01/arrayresponse.tex}}. \textcolor{black}{\input{actions01/do-we-use-theta.tex}}
We assume that the above reception, sampling and signal processing occur during a time interval termed the \ac{CPI} \cite{mishra2019toward}, which is an interval where sensing parameters (in our case $\theta_0$) remain unchanged. The received signal $\pmb{Y}_r \in \mathbb{C}^{N \times L}$ is the received radar vector and $\gamma_0$ is the complex channel gain of the reflected echo, containing two-way delay information between the \ac{DFRC} \textcolor{black}{\ac{BS}} and the intended target. The angle $\theta_0$ is the \ac{AoA} of the echo. Furthermore, due to the clutter present in the environment, the $n^{th}$ clutter source is assumed to be located at $\theta_n$ with complex amplitude $\gamma_n$. Also, $C$ is the number of clutter components in the environment. Moreover, similar to $\pmb{Z}_c$, the noise of the radar sub-system is i.i.d Gaussian modeled as $\pmb{Z}_r \sim \mathcal{N}(0,\sigma_r^2 \pmb{I}_N)$.

\textcolor{black}{\input{actions01/motivation-time-samples.tex}}
\subsection{Metrics}
To formulate a suitable and robust optimization problem that aims at solving \ac{ISAC} problems, it is crucial to define metrics related to the problem at hand. Based on \eqref{eq:y=Hxpz}, we can re-write the received signal as 
\begin{equation}
	\label{eq:received-signal-2}
	\pmb{Y}_c = \pmb{S} + \pmb{\MUI} + \pmb{Z}_c ,
\end{equation}
where $\pmb{\MUI} = \pmb{H}\pmb{X} - \pmb{S}$ is the multi-user interference and $\pmb{S} \in \mathbb{C}^{K \times L}$ is the desired signal carrying information symbols.

%


\textcolor{black}{\input{actions01/EMUI.tex}}

Furthermore, and from a radar perspective, we will consider a similarity constraint with a desired radar waveform, such as a chirp signal, \textcolor{black}{a \ac{LFM}} waveform, etc. Let the reference radar waveform be denoted as \textcolor{black}{$\pmb{x}_0$}, hence a similarity constraint could be expressed as a sphere centered at the desired waveform $\pmb{x}_0$ and with radius $\epsilon$, 
\begin{equation}
\label{eq:def-B-eps}
	\pmb{x} \in \mathcal{B}_{\epsilon}(\pmb{x}_0),
\end{equation}
where $\mathcal{B}_{\epsilon}(\pmb{x}_0) = \lbrace \pmb{x} , \Vert \pmb{x} - \pmb{x}_0 \Vert^2 \leq \epsilon^2 \rbrace$ \textcolor{black}{and $\pmb{x} = \ve(\pmb{X})$.}
Finally, the \ac{PAPR} is a waveform metric that shows the ratio of peak values to average power of that waveform. For example, a constant waveform enjoys a $\PAPR$ equal to one.    \textcolor{black}{To this end, the $\PAPR$ over an observation time of a signal vector $\pmb{x}$ of $NL$ samples, is given as:
\begin{equation}
	\PAPR(\pmb{x})
	=
	\frac{\max\limits_{\ell = 1 \ldots NL}\vert \pmb{x}(\ell) \vert^2}{\frac{1}{L}\sum\limits_{\ell = 1}^{NL} \vert \pmb{x}(\ell) \vert^2}.
\end{equation}
}
\vspace{-0.2cm}
\section{\ac{DFRC} Waveform Optimization Framework}
\label{sec:framework}
In this section, we formulate an optimization problem dedicated to maximizing the total achievable rate of communication users, with radar similarities and adjustable $\PAPR$. \textcolor{black}{Note that for fixed constellation energy, i.e. fixed $\mathbb{E}(\vert \pmb{S}_{k,\ell} \vert^2)$, minimizing the MUI energy can maximize the sum-rate \cite{liu2018toward,zhao2020mimo}. Based on this, we propose the following problem,}  
\begin{equation}
 \label{eq:problem1}
\begin{aligned}
(\mathcal{P}):
\begin{cases}
\min\limits_{\lbrace \pmb{x} \rbrace}&  \textcolor{black}{E_{\MUI}} \\
\textrm{s.t.}
 &  \PAPR(\pmb{x}) \leq \eta \\ 
 & \pmb{x} \in \mathcal{B}_{\epsilon}(\pmb{x}_0) ,\\ 
\end{cases}
\end{aligned}
\end{equation}
To enforce a unit power constraint over the transmitted waveforms, a total unit norm over all waveforms is integrated via a norm constraint, i.e. \textcolor{black}{$\Vert \pmb{x} \Vert^2 = 1$}. Combining this norm with the $\PAPR$ constraint, the above problem reads as 
\begin{equation}
 \label{eq:problem2}
\begin{aligned}
(\mathcal{P}):
\begin{cases}
\min\limits_{\lbrace \pmb{x} \rbrace}&  \textcolor{black}{E_{\MUI}} \\
\textrm{s.t.}
 &  \textcolor{black}{\Vert \pmb{x} \Vert^2 = 1,} \\ 
 & \pmb{x}^H \pmb{F}_n \pmb{x} \leq \frac{\eta}{NL} , \quad \forall n \\
 & \pmb{x} \in \mathcal{B}_{\epsilon}(\pmb{x}_0), \\ 
\end{cases}
\end{aligned}
\end{equation}
where $\pmb{F}_n$ is a matrix of all-zeros, except for $1$ located at its $n^{th}$ diagonal entry. \textcolor{black}{\input{actions01/following-set-of-logic.tex}}
\begin{equation}
 \label{eq:problem3}
\begin{aligned}
(\mathcal{P}'):
\begin{cases}
\min\limits_{\lbrace \pmb{x} \rbrace}& \textcolor{black}{ \Vert  \pmb{x} - \pmb{x}_{\comm} \Vert^2} \\
\textrm{s.t.}
 &  \textcolor{black}{\Vert \pmb{x} \Vert^2 = 1,} \\ 
 & \pmb{x}^H \pmb{F}_n \pmb{x} \leq \frac{\eta}{NL} , \quad \forall n \\
 & \pmb{x} \in \mathcal{B}_{\epsilon}(\pmb{x}_0), \\ 
\end{cases}
\end{aligned}
\end{equation}
It can be easily \textcolor{black}{verified} that the above problem is non-convex due to the norm-equality constraint.
\textcolor{black}{\input{actions01/precodingwaveform.tex}}
\section{\textcolor{black}{Impact of \ac{PAPR} on Sensing and Communications}}
\label{sec:impact}
\begin{figure}[t]
	\centering
	\includegraphics[width=1\linewidth]{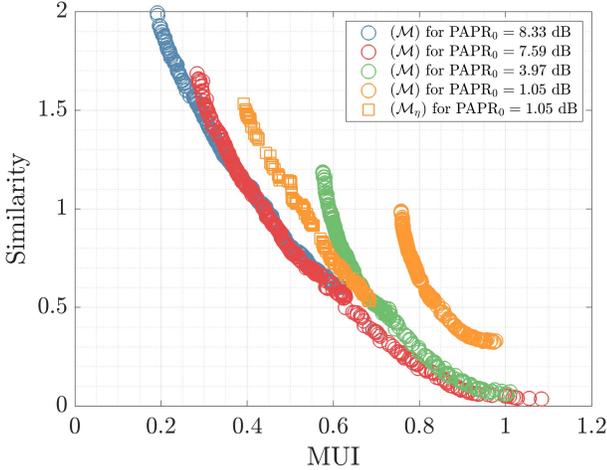}
	\caption{\textcolor{black}{Illustration of the Pareto boundaries of problems $(\mathcal{M})$ and $(\mathcal{M}_{\eta})$.}}
	\label{fig:Pareto}
\end{figure}
\textcolor{black}{\input{actions01/PAPR-MUI-Similarity.tex}}

\section{DFRC Waveform Design Via Alternating Directions method of multipliers}
\label{sec:admm}
Before we proceed, we convert the problem to real-valued vectors as follows
\begin{equation}
 \label{eq:problem3}
\begin{aligned}
(\bar{\mathcal{P}}'):
\begin{cases}
\min\limits_{\lbrace \bar{\pmb{x}} \rbrace}&  \textcolor{black}{\Vert   \bar{\pmb{x}} -  \bar{\pmb{x}}_{\comm} \Vert^2} \\
\textrm{s.t.}
 &  \textcolor{black}{\Vert \bar{\pmb{x}} \Vert^2 = 1,} \\ 
 &  \bar{\pmb{x}}^H \bar{\pmb{F}}_n  \bar{\pmb{x}} \leq \frac{\eta}{NL} , \quad \forall n \\
 &  \bar{\pmb{x}} \in \mathcal{B}_{\epsilon}( \bar{\pmb{x}}_0), \\ 
\end{cases}
\end{aligned}
\end{equation}
where $\bar{\pmb{x}} = \begin{bmatrix} \Real(\pmb{x})^T  &  \Imag(\pmb{x})^T \end{bmatrix}^T$ and the same definition applies to $\bar{\pmb{x}}_0$ and $\bar{\pmb{x}}_{\comm}$.  Furthermore, matrix $\bar{\pmb{F}}_n \in \mathbb{R} ^{2NL \times 2NL}$ is an all-zero matrix except its $n^{th}$ and $(NL+n)^{th}$ entries that are set to $1$. Next, introducing auxiliary variables $\pmb{\alpha}, \pmb{\beta}, \pmb{\gamma}_1 \ldots \pmb{\gamma}_{NL}$ to the problem reads,
\begin{equation}
 \label{eq:final-problem}
\begin{aligned}
(\bar{\mathcal{P}}'):
\begin{cases}
\min\limits_{\lbrace \bar{\pmb{x}} \rbrace}&  \textcolor{black}{\Vert   \bar{\pmb{x}} -  \bar{\pmb{x}}_{\comm} \Vert^2} \\
\textrm{s.t.}
  & \pmb{\alpha}  = \bar{\pmb{x}}, \quad \pmb{\alpha}^T \pmb{\alpha} = 1,  \\
  & \pmb{\beta} = \bar{\pmb{x}} - \pmb{x}_0, \quad \pmb{\beta}^T \pmb{\beta} \leq \epsilon^2 ,   \\
  & \pmb{\gamma}_n  = \bar{\pmb{F}}_n  \bar{\pmb{x}} , \quad  \pmb{\gamma}_n^T \pmb{\gamma}_n  \leq \frac{\eta}{NL} , \quad \forall n \\
\end{cases}
\end{aligned}
\end{equation}
The augmented Lagrangian of the above problem is 
\begin{equation}
\label{eq:augmented-lag-multiplier-all}
\begin{split}
	&\mathcal{L}_{\rho}(\bar{\pmb{x}},\pmb{\alpha},\pmb{\beta},\textcolor{black}{\pmb{\gamma}},\pmb{u},\pmb{v},\textcolor{black}{\pmb{w}}) \\
	&=
	\textcolor{black}{\Vert \bar{\pmb{x}} -  \bar{\pmb{x}}_{\comm} \Vert^2}
	+ \pmb{u}^T ( \bar{\pmb{x}} - \pmb{\alpha} ) + \pmb{v}^T ( \bar{\pmb{x}} - \bar{\pmb{x}}_0 - \pmb{\beta}) \\
	&+  \sum\limits_{n=1}^{NL} \pmb{w}_n^T (\bar{\pmb{F}}_n  \bar{\pmb{x}}  - \pmb{\gamma}_n) + \frac{\rho}{2} \Vert \bar{\pmb{x}} - \pmb{\alpha} \Vert^2 + \frac{\rho}{2} \Vert \bar{\pmb{x}} - \bar{\pmb{x}}_0 - \pmb{\beta} \Vert^2 \\
	&+  \frac{\rho}{2} \sum\limits_{n=1}^{NL} \Vert \bar{\pmb{F}}_n  \bar{\pmb{x}}  - \pmb{\gamma}_n \Vert^2, 
\end{split}
\end{equation}
where \textcolor{black}{$\pmb{\gamma} = \begin{bmatrix} \pmb{\gamma}_1^T  \ldots \pmb{\gamma}_{NL}^T \end{bmatrix}^T $, $\pmb{w} = \begin{bmatrix} \pmb{w}_1^T  \ldots \pmb{w}_{NL}^T \end{bmatrix}^T $} and $\rho > 0$ is a penalty parameter. It is worth noting that this augmented Lagrangian function could be thought of as the non-augmented Lagrangian of the following optimization problem  
\begin{equation}
 \label{eq:final-problem-equivalent-nonaugmented}
\begin{aligned}
(\bar{\mathcal{P}}_\rho'):
\begin{cases}
\min\limits_{\lbrace \bar{\pmb{x}} \rbrace}&  \textcolor{black}{\Vert   \bar{\pmb{x}} -  \bar{\pmb{x}}_{\comm} \Vert^2} 
+ \frac{\rho}{2} \Vert \bar{\pmb{x}} - \pmb{\alpha} \Vert^2 \\ &
+ \frac{\rho}{2} \Vert \bar{\pmb{x}} - \bar{\pmb{x}}_0 - \pmb{\beta} \Vert^2 
+  \frac{\rho}{2} \sum\limits_{n=1}^{NL} \Vert \bar{\pmb{F}}_n  \bar{\pmb{x}}  - \pmb{\gamma}_n \Vert^2 \\
\textrm{s.t.}
  & \pmb{\alpha}  = \bar{\pmb{x}}, \quad  \pmb{\beta} = \bar{\pmb{x}} - \pmb{x}_0,  \quad  \pmb{\gamma}_n  = \bar{\pmb{F}}_n  \bar{\pmb{x}} , \quad \forall n \\
\end{cases}
\end{aligned}
\end{equation}
The variables are updated in a round-robin fashion as follows
\textcolor{black}{\input{actions01/admm-round-robin-equations.tex}}
where sets $\mathcal{C}_{\alpha}, \mathcal{C}_{\beta}, \mathcal{C}_{\gamma}$ are defined as follows
\begin{subequations}
	\begin{equation}
		\mathcal{C}_{\alpha} = \lbrace \pmb{\alpha} , \quad \pmb{\alpha}^T \pmb{\alpha} = 1 	 \rbrace , 
	\end{equation}
	\begin{equation}
		\mathcal{C}_{\beta} = \lbrace \pmb{\beta} , \quad  \pmb{\beta}^T \pmb{\beta} \leq \epsilon^2  \rbrace , 
	\end{equation}
	\begin{equation}
		\mathcal{C}_{\gamma} = \lbrace \pmb{\gamma} , \quad \pmb{\gamma}^T \pmb{\gamma} \leq \frac{\eta}{NL} \rbrace , 
	\end{equation}
\end{subequations}
At the $(m+1)^{th}$ iteration, we update $\bar{\pmb{x}}^{(m+1)}$ by solving
\begin{equation}
\label{eq:first-opt}
	\bar{\pmb{x}}^{(m+1)}
	=
	\arg\min_{\bar{\pmb{x}}}
	\mathcal{L}_{\rho}(\bar{\pmb{x}},\pmb{\alpha}^{(m)},\pmb{\beta}^{(m)},\textcolor{black}{\pmb{\gamma}^{(m)}},\pmb{u}^{(m)},\pmb{v}^{(m)},\textcolor{black}{\pmb{w}^{(m)}}).
\end{equation}
Setting the gradient of $\mathcal{L}_{\rho}$ with respect to $\bar{\pmb{x}}$, to zero, i.e.
\begin{equation}
\label{eq:gradient-at-x}
	\nabla_{\bar{\pmb{x}}}
	\Big[
	\mathcal{L}_{\rho}(\bar{\pmb{x}},\pmb{\alpha}^{(m)},\pmb{\beta}^{(m)},\textcolor{black}{\pmb{\gamma}^{(m)}},\pmb{u}^{(m)},\pmb{v}^{(m)},\textcolor{black}{\pmb{w}^{(m)}})
	\Big]_{\bar{\pmb{x}} = \bar{\pmb{x}}^{(m+1)} } 
	=
	\pmb{0},
\end{equation}
we get 
\begin{equation}
\begin{split}
\label{eq:gradient-at-x-expanded}
	2( \textcolor{black}{\bar{\pmb{x}}^{(m+1)}} &- \textcolor{black}{\bar{\pmb{x}}_{\comm}} ) + \pmb{u}^{(m)} + \pmb{v}^{(m)} + \sum\limits_{n=1}^{NL} \bar{\pmb{F}}_n^T \pmb{w}_n^{(m)} \\
	&+ \rho (\bar{\pmb{x}}^{(m+1)} - \pmb{\alpha}^{(m)}) 
	+ \rho(  \bar{\pmb{x}}^{(m+1)} - \bar{\pmb{x}}_0 - \pmb{\beta}^{(m)} )\\
	&+ \rho \sum\limits_{n=1}^{NL} \bar{\pmb{F}}_n^T(\bar{\pmb{F}}_n \bar{\pmb{x}}^{(m+1)} - \pmb{\gamma}_n^{(m)})  = \pmb{0}.
\end{split}
\end{equation}
Using the symmetric and idempotent property of $\bar{\pmb{F}}_n$, i.e. $\bar{\pmb{F}}_n^T = \bar{\pmb{F}}_n$ and $\bar{\pmb{F}}_n\bar{\pmb{F}}_n=\bar{\pmb{F}}_n$, we can arrange the above gradient expression to get an update equation on \textcolor{black}{$\bar{\pmb{x}}^{(m+1)}$} as follows,
\begin{equation}
\label{eq:x-update}
\boxed{
\begin{split}
	\bar{\pmb{x}}^{(m+1)} 
	&=
	\frac{1}{2+3\rho}
	\Bigg(
	2 \bar{\pmb{x}}_{\comm} 
	- \pmb{u}^{(m)} 
	- \pmb{v}^{(m)}
	- \sum\limits_{n=1}^{NL}\bar{\pmb{F}}_n \pmb{w}_n^{(m)} \\ &
	+ \rho \pmb{\alpha}^{(m)} + \rho (\bar{\pmb{x}}_0 + \pmb{\beta}^{(m)} )  
	+ \rho \sum\limits_{n=1}^{NL} \bar{\pmb{F}}_n \pmb{\gamma}_n^{(m)}
	\Bigg).
\end{split}
}
\end{equation}
Now that we have an updated waveform at iteration $(m+1)$, we can proceed to update auxiliary variable $\pmb{\alpha}^{(m+1)}$,
\begin{equation}
\label{eq:second-opt}
	\arg\min_{\pmb{\alpha} \in \mathcal{C}_{\alpha}}
	\mathcal{L}_{\rho}(\bar{\pmb{x}}^{(m+1)},\pmb{\alpha},\pmb{\beta}^{(m)},\textcolor{black}{\pmb{\gamma}^{(m)}},\pmb{u}^{(m)},\pmb{v}^{(m)},\textcolor{black}{\pmb{w}^{(m)}}),
\end{equation}
which could be formulated as an optimization problem in $\pmb{\alpha}$, 
\begin{equation}
 \label{eq:update-alpha-problem}
\begin{aligned}
(\bar{\mathcal{P}}_\alpha):
\begin{cases}
\min\limits_{\lbrace \pmb{\alpha} \rbrace}&  \mathcal{L}_{\rho}(\bar{\pmb{x}}^{(m+1)},\pmb{\alpha},\pmb{\beta}^{(m)},\textcolor{black}{\pmb{\gamma}^{(m)}},\pmb{u}^{(m)},\pmb{v}^{(m)},\textcolor{black}{\pmb{w}^{(m)}})\\
\textrm{s.t.}
 &  \pmb{\alpha} \in \mathcal{C}_\alpha.
\end{cases}
\end{aligned}
\end{equation}
By \textcolor{black}{omitting} terms that are independent of $\pmb{\alpha}$, the optimization problem in equation \eqref{eq:update-alpha-problem} can be equivalently expressed as 
\begin{equation}
 \label{eq:update-alpha-problem-equivalent}
\begin{aligned}
(\bar{\mathcal{P}}_\alpha):
\begin{cases}
\min\limits_{\lbrace \pmb{\alpha} \rbrace}&  (\pmb{u}^{(m)})^T ( \bar{\pmb{x}}^{(m+1)} - \pmb{\alpha} ) + \frac{\rho}{2} \Vert \bar{\pmb{x}}^{(m+1)} - \pmb{\alpha} \Vert^2\\
\textrm{s.t.}
 &  \pmb{\alpha}^T\pmb{\alpha} = 1,
\end{cases}
\end{aligned}
\end{equation}
whose Lagrangian will be denoted as $\mathcal{L}_\alpha$ and is given as 
\begin{equation}
\begin{split}
	\mathcal{L}_\alpha(\pmb{\alpha},\lambda)  
	&=
	(\pmb{u}^{(m)})^T ( \bar{\pmb{x}}^{(m+1)} - \pmb{\alpha} ) \\ & + \frac{\rho}{2} \Vert \bar{\pmb{x}}^{(m+1)} - \pmb{\alpha} \Vert^2
	- \lambda ( \pmb{\alpha}^T\pmb{\alpha} - 1 ).
\end{split}
\end{equation}
Setting the gradient with respect to $\pmb{\alpha}$ to zero,
\begin{equation}
	\nabla_{\pmb{\alpha}}
	\Big[
	\mathcal{L}_\alpha(\pmb{\alpha},\lambda)
	\Big]_{\pmb{\alpha} = \pmb{\alpha}^{(m+1)}}
	=
	\pmb{0},
\end{equation}
we get
\begin{equation}
	-\pmb{u}^{(m)}
	+\rho(\pmb{\alpha}^{(m+1)} - \bar{\pmb{x}}^{(m+1)}) - 2 \lambda \pmb{\alpha}^{(m+1)} = \pmb{0},
\end{equation}
which is further expressed as 
\begin{equation}
	\pmb{\alpha}^{(m+1)}
	=
	\frac{1}{\rho - 2 \lambda}\Big(
	\pmb{u}^{(m)} + \rho \bar{\pmb{x}}^{(m+1)}
	\Big).
\end{equation}
Without loss of generality, the Lagrangian multiplier $\lambda$ can be set to $0$ and the vector $\pmb{\alpha}^{(m+1)}$ can be normalized to satisfy the constraint $(\pmb{\alpha}^{(m+1)})^T \pmb{\alpha}^{(m+1)} = 1$, as follows
\begin{equation}
\label{eq:alpha-update}
\boxed{
	\pmb{\alpha}^{(m+1)}
	=
	\frac{\bar{\pmb{x}}^{(m+1)} + \frac{1}{\rho} \pmb{u}^{(m)} }{\Vert \bar{\pmb{x}}^{(m+1)} + \frac{1}{\rho} \pmb{u}^{(m)}  \Vert }.
	}
\end{equation}
Now that we have an updated value of both $\pmb{x}^{(m+1)}$ and $\pmb{\alpha}^{(m+1)}$, we can go ahead and update $\pmb{\beta}$ by solving 
\begin{equation}
\label{eq:third-opt}
	\arg\min_{\pmb{\beta} \in \mathcal{C}_{\beta}}
	\mathcal{L}_{\rho}(\bar{\pmb{x}}^{(m+1)},\pmb{\alpha}^{(m+1)},\pmb{\beta},\textcolor{black}{\pmb{\gamma}^{(m)}},\pmb{u}^{(m)},\pmb{v}^{(m)},\textcolor{black}{\pmb{w}^{(m)}})
\end{equation}
In a very similar way, we can formulate the following optimization problem, 
\begin{equation}
 \label{eq:update-beta-problem}
\begin{aligned}
(\bar{\mathcal{P}}_\beta):
\begin{cases}
\min\limits_{\lbrace \pmb{\beta} \rbrace}&  \mathcal{L}_{\rho}(\bar{\pmb{x}}^{(m+1)},\pmb{\alpha}^{(m+1)},\pmb{\beta},\textcolor{black}{\pmb{\gamma}^{(m)}},\pmb{u}^{(m)},\pmb{v}^{(m)},\textcolor{black}{\pmb{w}^{(m)}})\\
\textrm{s.t.}
 &  \pmb{\beta} \in \mathcal{C}_\beta.
\end{cases}
\end{aligned}
\end{equation}
Ignoring terms that do not depend on $\pmb{\beta}$, the optimization problem in \eqref{eq:update-beta-problem} boils down to
\begin{equation}
 \label{eq:update-beta-problem-equivalent}
\begin{aligned}
(\bar{\mathcal{P}}_\beta):
\begin{cases}
\min\limits_{\lbrace \pmb{\beta} \rbrace}&  (\pmb{v}^{(m)})^T ( \bar{\pmb{x}}^{(m+1)} - \bar{\pmb{x}}_0 - \pmb{\beta})  \\ &  \qquad + \frac{\rho}{2} \Vert \bar{\pmb{x}}^{(m+1)} - \bar{\pmb{x}}_0 - \pmb{\beta} \Vert^2\\
\textrm{s.t.}
 &  \pmb{\beta}^T \pmb{\beta} \leq \epsilon^2.
\end{cases}
\end{aligned}
\end{equation}
To this end, the solution of the above problem is given as 
\begin{equation}
 \label{eq:beta-update}
\boxed{
	\pmb{\beta}^{(m+1)}
	=
\begin{cases}
\bar{\pmb{x}}^{(m+1)} - \bar{\pmb{x}}_0 + \frac{1}{\rho} \pmb{v}^{(m)} , & \text{if  } \in \mathcal{C}_\beta \\
		\epsilon
	\frac{\bar{\pmb{x}}^{(m+1)} - \bar{\pmb{x}}_0 + \frac{1}{\rho} \pmb{v}^{(m)} }{\Vert \bar{\pmb{x}}^{(m+1)} - \bar{\pmb{x}}_0  + \frac{1}{\rho} \pmb{v}^{(m)}  \Vert } , & \text{otherwise} .
\end{cases}
	}
\end{equation}
\textcolor{black}{\input{actions01/text-below-beta.tex}}
\begin{equation}
\label{eq:fourth-opt}
	\arg\min_{\textcolor{black}{\lbrace \pmb{\gamma}_n \in \mathcal{C}_{\gamma} \rbrace_{n=1}^{NL}}}
	\mathcal{L}_{\rho}(\textcolor{black}{\begin{bmatrix}
		\bar{\pmb{x}},\pmb{\alpha},\pmb{\beta}
	\end{bmatrix}^{(m+1)}},\textcolor{black}{\pmb{\gamma}},\pmb{u}^{(m)},\pmb{v}^{(m)},\textcolor{black}{\pmb{w}^{(m)}}),
\end{equation}
where the above minimization problem can be formulated as \textcolor{black}{$NL$ independent optimization problems}
\begin{equation}
 \label{eq:update-gamma-problem-equivalent}
\begin{aligned}
(\bar{\mathcal{P}}_\gamma):
\begin{cases}
\min\limits_{\lbrace \pmb{\gamma}_n \rbrace}&  (\pmb{w}_n^{(m)})^T (\bar{\pmb{F}}_n  \bar{\pmb{x}}^{(m+1)}  - \pmb{\gamma}_n) \\ & \qquad + \frac{\rho}{2} \Vert \bar{\pmb{F}}_n  \bar{\pmb{x}}^{(m+1)}  - \pmb{\gamma}_n \Vert^2 \\
\textrm{s.t.}
 &  \textcolor{black}{\pmb{\gamma}_n} \in \mathcal{C}_\gamma,
\end{cases}
\end{aligned}
\end{equation}
\textcolor{black}{for $n = 1\ldots NL$.} Note that in the above, we have removed terms that do not depend on $\pmb{\gamma}_n$. The solution of $(\bar{\mathcal{P}}_\gamma)$ in equation \eqref{eq:update-gamma-problem-equivalent} is similar to the solution appearing in equation \eqref{eq:beta-update} and is given as 
\begin{equation}
\label{eq:gamma-update}
\boxed{
	\pmb{\gamma}_n^{(m+1)}  
	=
	\begin{cases}
	\bar{\pmb{F}}_n  \bar{\pmb{x}}^{(m+1)} + \frac{1}{\rho} \pmb{w}_n^{(m)}, & \text{if }  \in \mathcal{C}_\gamma \\
		\sqrt{\frac{\eta}{NL}}\frac{ \bar{\pmb{F}}_n  \bar{\pmb{x}}^{(m+1)} + \frac{1}{\rho} \pmb{w}_n^{(m)} }{\Vert \bar{\pmb{F}}_n  \bar{\pmb{x}}^{(m+1)} + \frac{1}{\rho} \pmb{w}_n^{(m)} \Vert},
		& \text{otherwise}.
	\end{cases}
	}
\end{equation}
Then the auxiliary variables are updated according to \ac{ADMM}'s dual variable update of step size $\rho$, which is the same as the augmented Lagrangian parameter \cite{boyd2011distributed}, i.e.
\begin{align}
	\pmb{u}^{(m+1)} &= \pmb{u}^{(m)} + \rho(\bar{\pmb{x}}^{(m+1)} - \pmb{\alpha}^{(m+1)}), \label{eq:u-update} \\
	\pmb{v}^{(m+1)} &= \pmb{v}^{(m)} + \rho(\bar{\pmb{x}}^{(m+1)} - \pmb{x}_0 - \pmb{\beta}^{(m+1)}), \label{eq:v-update}  \\
	\pmb{w}_n^{(m+1)} &= \pmb{w}_n^{(m)} + \rho(\bar{\pmb{F}}_n  \bar{\pmb{x}}^{(m+1)} - \pmb{\gamma}_n^{(m+1)}), \textcolor{black}{\quad \forall n.} \label{eq:w-update}
\end{align}
Note that the above update involves all the most recent quantity values. A summary of the proposed \ac{ADMM}-based \ac{DFRC} waveform design is summarized in \textcolor{black}{\textbf{Algorithm \ref{alg:cap}}}.

\textcolor{black}{\input{actions01/algorithm1.tex}}

\section{Convergence Analysis}
\label{sec:convergence}
For convergence analysis, it is more convenient to express the augmented Lagrangian function as follows
\begin{equation}
\label{eq:another-expression-of-Lrho}
	\mathcal{L}_\rho
	(\bar{\pmb{x}},\pmb{\lambda},\pmb{\mu})
	=
	f(\bar{\pmb{x}}) + \pmb{\lambda}^T \pmb{e} + \frac{\rho}{2} \Vert \pmb{e} \Vert^2, 
\end{equation}
with $\pmb{e} = \pmb{A}\bar{\pmb{x}} - \pmb{\mu} - \pmb{c} $ is a measure of drift (or feasibility gap) that reflects feasibility of the constraints and $\pmb{\lambda} = \begin{bmatrix} 
\pmb{u}^T & \pmb{v}^T & \pmb{w}_1^T  & \ldots & \pmb{w}_{NL}^T \end{bmatrix}^T$ contains all Lagrangian variables and $\pmb{\mu} = \begin{bmatrix} 
\pmb{\alpha}^T & \pmb{\beta}^T & \pmb{\gamma}_1^T  & \ldots & \pmb{\gamma}_{NL}^T \end{bmatrix}^T$ contains all the auxiliary variables. Furthermore, $\pmb{A} = \begin{bmatrix}
	\pmb{1}_2 \otimes \pmb{I}_{2NL} \\ \pmb{F}
\end{bmatrix}$ and $\pmb{F}$ is a block-matrix stacking all matrices $\pmb{F}_n$ one on top of the other. Moreover, $\pmb{c} = \pmb{o}_2 \otimes \bar{\pmb{x}}_0$  \textcolor{black}{and $f(\bar{\pmb{x}}) = \Vert \bar{\pmb{x}} - \bar{\pmb{x}}_{\comm} \Vert^2$. }  \\\\
\textbf{Lemma 1} \textit{( \textcolor{black}{$\bar{\pmb{x}}^{(m+1)}$ minimizes $g^{(m+1)}(\pmb{x})$} ): \\ Let $g$ be defined as follows}
\textcolor{black}{
\begin{equation}
\label{eq:alternative-expression}
	g^{(m+1)}(\pmb{x}) = f(\pmb{x}) 
	 + \pmb{x}^T \pmb{A}^T \Big(\pmb{\lambda}^{(m+1)} + \rho  (\pmb{\mu}^{(m+1)} - \pmb{\mu}^{(m)}) \Big),
\end{equation}
}
\textit{then $\bar{\pmb{x}}^{(m+1)}$ is its minimizer.}\\
\textbf{Proof} See Appendix \ref{lemma1-proof}.  \\\\
\textbf{Lemma 2} \textit{($\pmb{\mu}^{(m+1)}$ minimizes $h^{(m+1)}(\pmb{\mu})$): \\ Let $h$ be defined as follows}
\begin{equation}
	h^{(m+1)}(\pmb{\mu}) = - \pmb{\mu}^T  \pmb{\lambda}^{(m+1)} ,
\end{equation}
\textit{then \textcolor{black}{${\pmb{\mu}}^{(m+1)}$} is its minimizer.}\\
\textbf{Proof} See Appendix \ref{lemma2-proof}.\\\\
Next, we present a lemma that re-expresses the inner product of the difference between $\pmb{\lambda}^{(m+1)}$ and an arbitrary point $\pmb{\lambda}$ onto the residual part at the $(m+1)^{th}$ iteration, i.e. $\pmb{e}^{(m+1)}$. The resulting expression turns out to be more suitable for convergence analysis, as the resulting expression contains norm typed quantities. The lemma is given as follows\\\\ 
\textbf{Lemma 3} \textit{At the $(m+1)^{th}$ iteration of \textcolor{black}{\textbf{Algorithm \ref{alg:cap}}}, the following equality holds true}
\begin{equation}
\begin{split}
& (\pmb{\lambda}^{(m+1)} - \pmb{\lambda})^T \pmb{e}^{(m+1)}
	\\ & =
	\frac{1}{2\rho}(\Vert \pmb{\lambda}^{(m+1)} - \pmb{\lambda} \Vert^2  - \Vert \pmb{\lambda}^{(m)} - \pmb{\lambda} \Vert^2) + \frac{\rho}{2} \Vert \pmb{e}^{(m+1)} \Vert^2.
\end{split}
\end{equation}
\textbf{Proof} See Appendix \ref{lemma3-proof}.  \\\\
Before revealing a convergence property of the \ac{ADMM}-based \ac{DFRC} waveform design described in \textcolor{black}{\textbf{Algorithm \ref{alg:cap}}}, we introduce a direct consequence of the previous lemmas, which reveals a decreasing behaviour of the auxiliary variables when projected onto the residual at the $(m+1)^{th}$ iteration. \\\\
\textbf{Consequence 1} \textit{At the $(m+1)^{th}$ iteration of \textcolor{black}{\textbf{Algorithm \ref{alg:cap}}}, we have the following inequality}
\begin{equation}
	 - (\pmb{ \mu }^{(m+1)})^T  \pmb{e}^{(m+1)} 
	 \leq
	 - (\pmb{ \mu }^{(m)})^T  \pmb{e}^{(m+1)}.
\end{equation}
\textbf{Proof} See Appendix \ref{consequence1-proof}.  \\\\
We can now formulate a theorem that describes the convergence behaviour of the proposed \ac{ADMM}-based \ac{DFRC} waveform design iterative method described in \textcolor{black}{\textbf{Algorithm \ref{alg:cap}}}. To this end, we have the following \\\\
\textbf{Theorem 1} \textit{Consider the iterative method presented in \textcolor{black}{\textbf{Algorithm \ref{alg:cap}}}. Regardless of initialization, the method is guaranteed convergence in the following sense }
\begin{equation}
	\Vert \pmb{e}^{(m)} \Vert^2 \rightarrow 0 ,
\end{equation}
and
\begin{equation}
	\Vert \pmb{\mu}^{(m+1)} - \pmb{\mu}^{(m)} \Vert^2 \rightarrow 0 .
\end{equation}
\textbf{Proof} See Appendix \ref{theorem1-proof}.

\section{\textcolor{black}{Computational Complexity Analysis}}
\label{sec:complexity}
\textcolor{black}{\input{actions01/complexity.tex}}

\section{\textcolor{black}{Imperfect \ac{CSI} Extension}}
\label{sec:imperfect-CSI} 
\textcolor{black}{\input{actions01/imperfect_CSI.tex}}

\begin{figure*}[!t]
\centering
\subfloat[]{\includegraphics[width=3.5in]{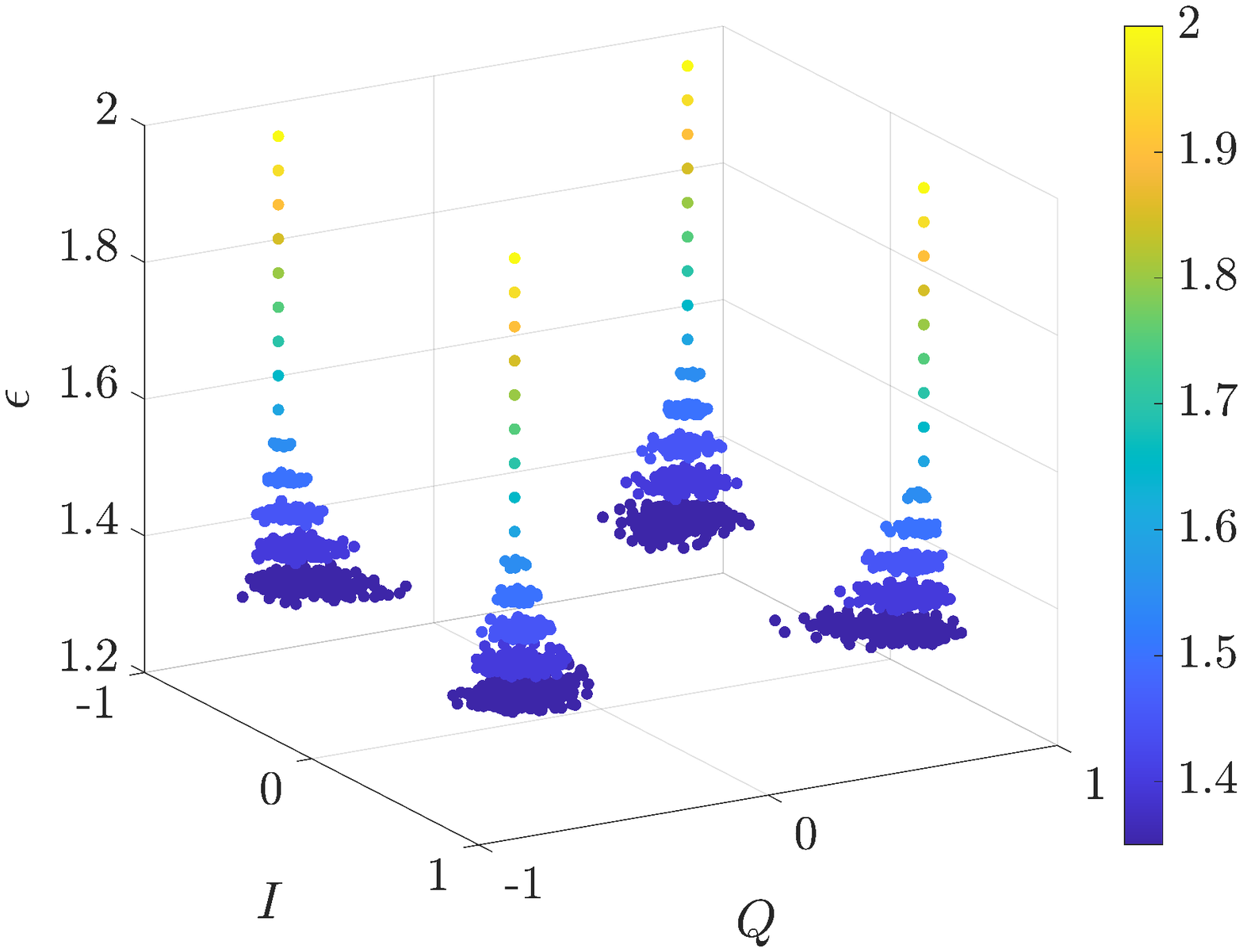}%
\label{fig_first_case}}
\hfil
\subfloat[]{\includegraphics[width=3.5in]{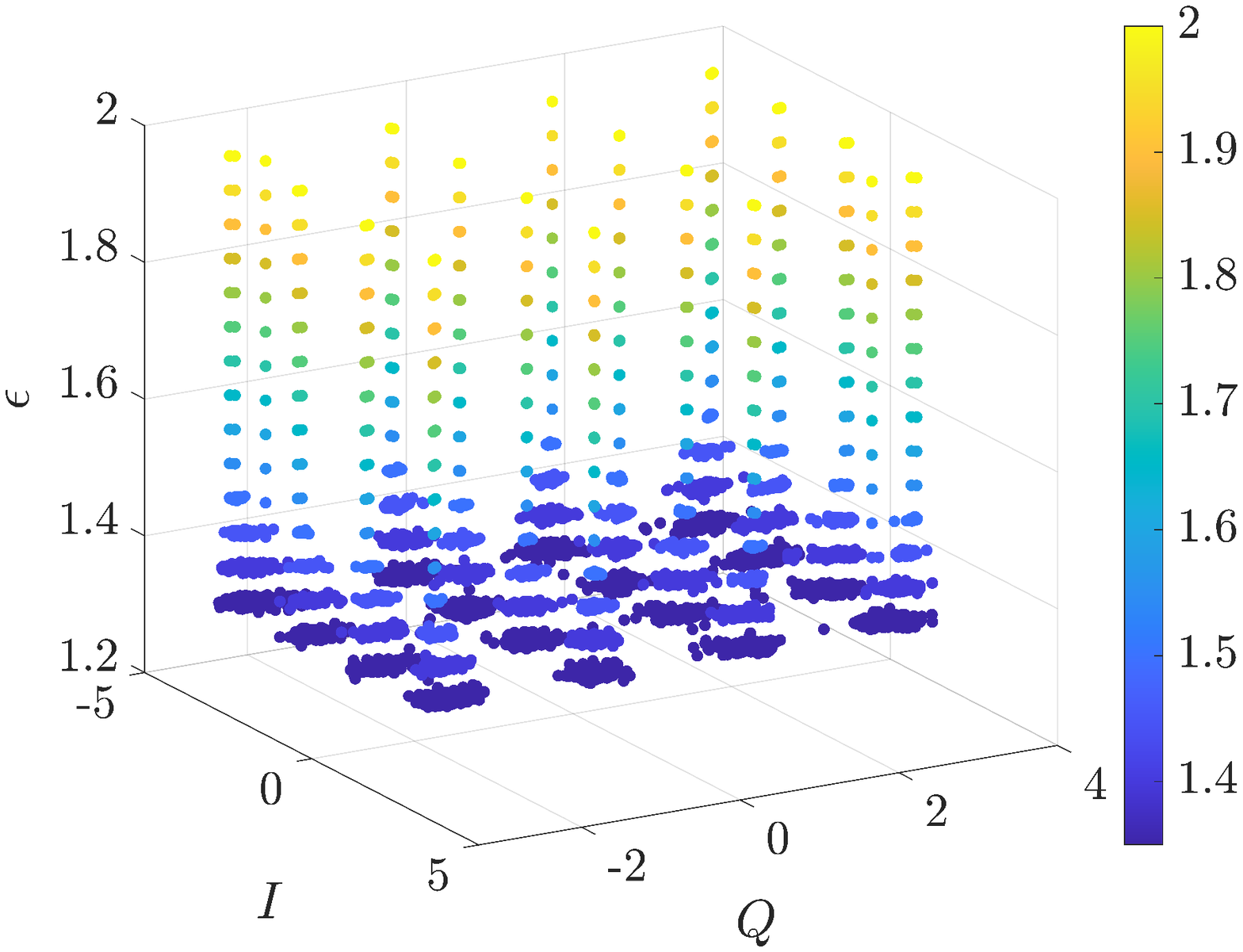}%
\label{fig_second_case}}
\caption{The constellation produced by the \ac{ADMM}-based \ac{DFRC} waveform design and seen at the receiver, as a function of different values of $\epsilon$ for (a) QPSK (b) $16-$QAM.}
\label{fig:3DConstellations}
\end{figure*}

\begin{figure}[t]		
	\centering
	\includegraphics[width=1\linewidth]{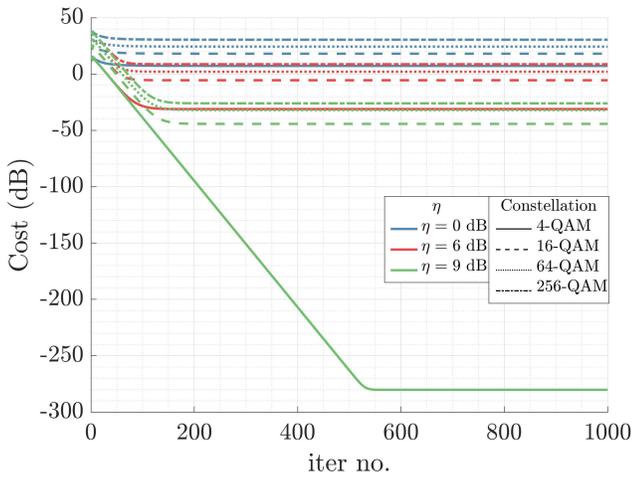}
	\caption{\textcolor{black}{The convergence behavior of \textcolor{black}{\textbf{Algorithm \ref{alg:cap}}} for different values of $\eta$ and $\epsilon = 1.85$, where each line is averaged over Monte Carlo trials.}}
	\label{fig:CostvsIter}
\end{figure}

\begin{figure}[t]
	\centering
	\includegraphics[width=1\linewidth]{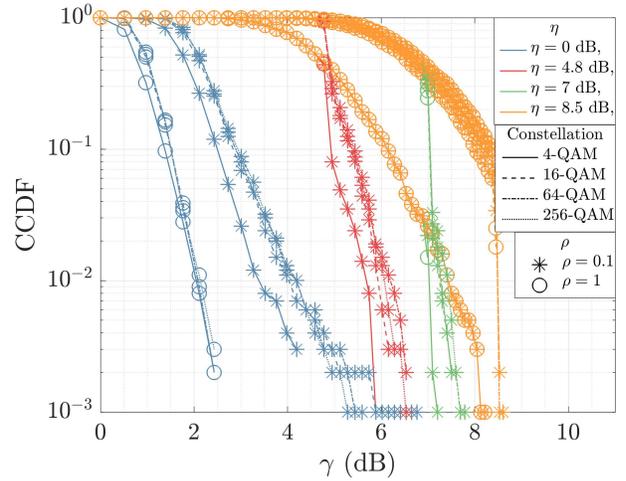}
	\caption{\textcolor{black}{The CCDF of the generated waveform described in \textbf{Algorithm \ref{alg:cap}} for different values of $\eta$, $\rho$ and constellation sizes.}}
	\label{fig:CCDF}
\end{figure}

\begin{figure}[t]
	\centering
	\includegraphics[width=1\linewidth]{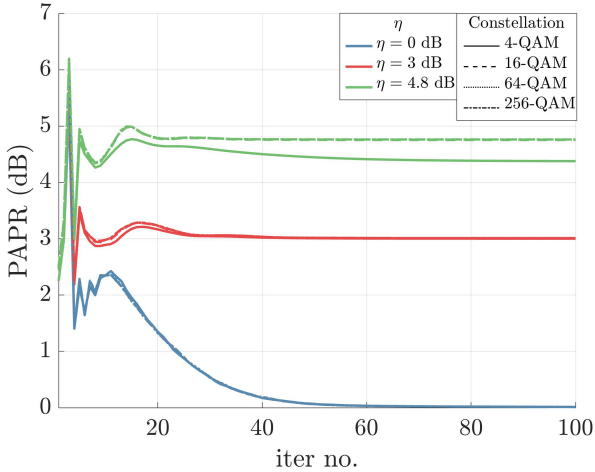}
	\caption{\textcolor{black}{The convergence behavior of the \ac{PAPR} per iteration of \textcolor{black}{\textbf{Algorithm \ref{alg:cap}}} for different values of $\eta$ , where each line is averaged over Monte Carlo trials.}}
	\label{fig:PAPRvsIter}
\end{figure}

\begin{figure}[t]
	\centering
	\includegraphics[width=1\linewidth]{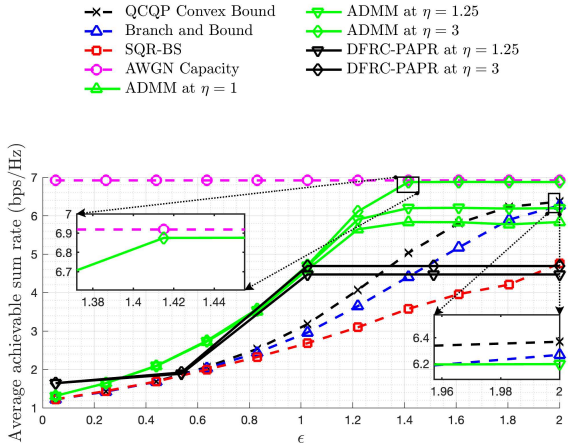}
	\caption{\textcolor{black}{Trade-off between the communication average achievable sum- rate and similarity waveform radar measure at $\SNR=10$ dB for $N = 4$ antennas and $K = 2$ communication users. The constellation utilized is QPSK.}}
	\label{fig:SumRate}
\end{figure}
\begin{figure*}[!t]
\centering
\subfloat[]{\includegraphics[width=2in]{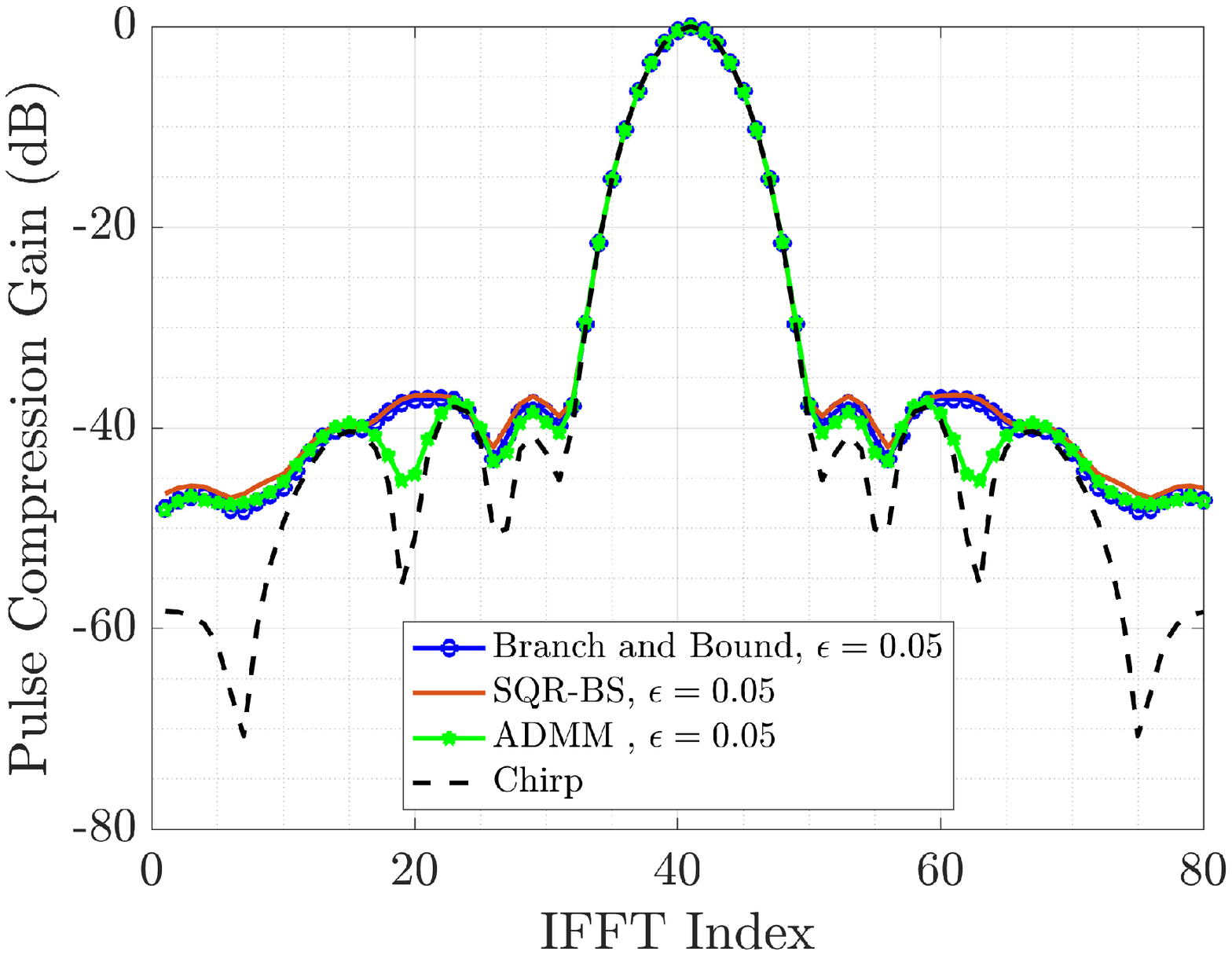}%
\label{fig:PulseGainCompression_a}}
\hfil
\subfloat[]{\includegraphics[width=2in]{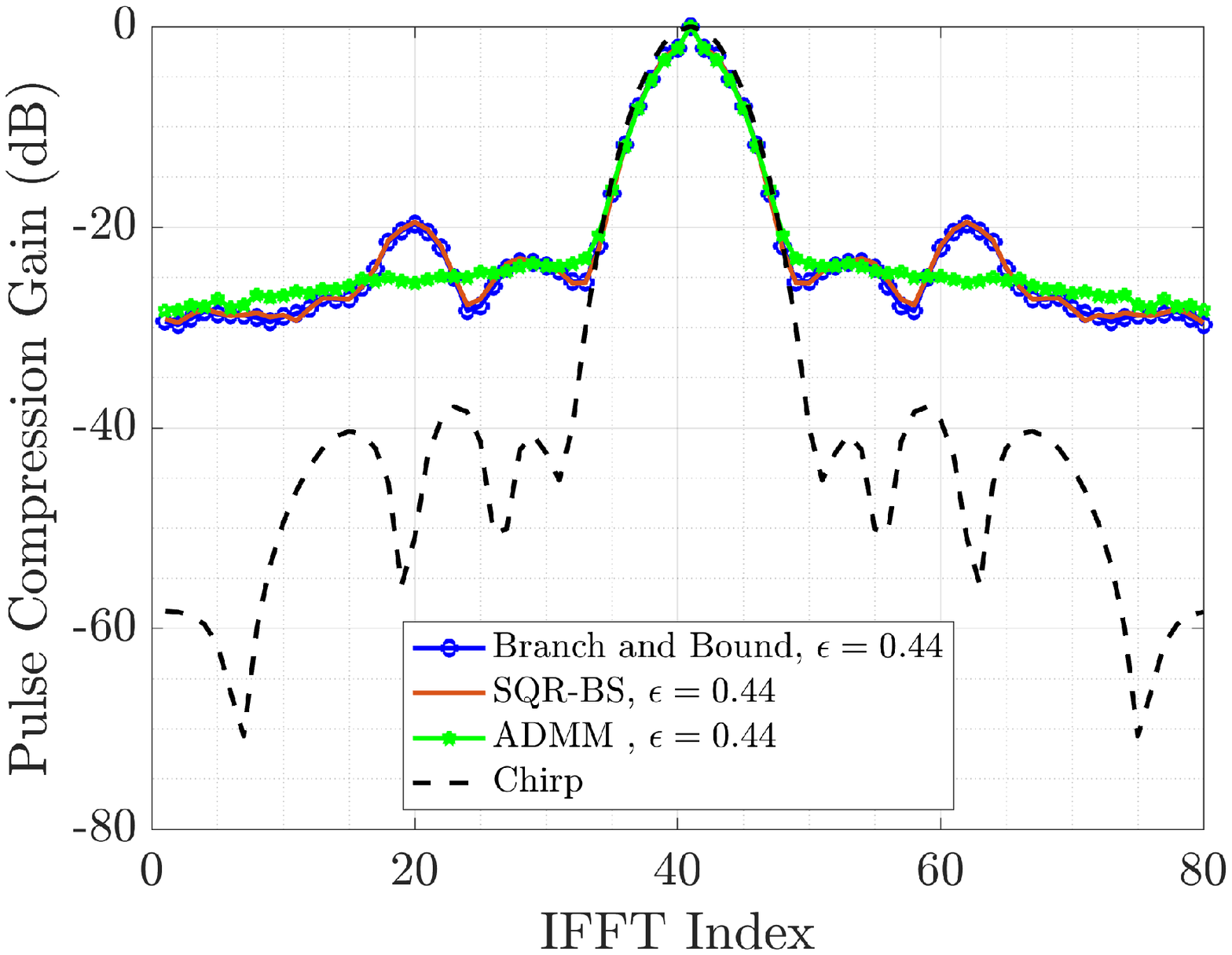}%
\label{fig:PulseGainCompression_b}}
\hfil
\subfloat[]{\includegraphics[width=2in]{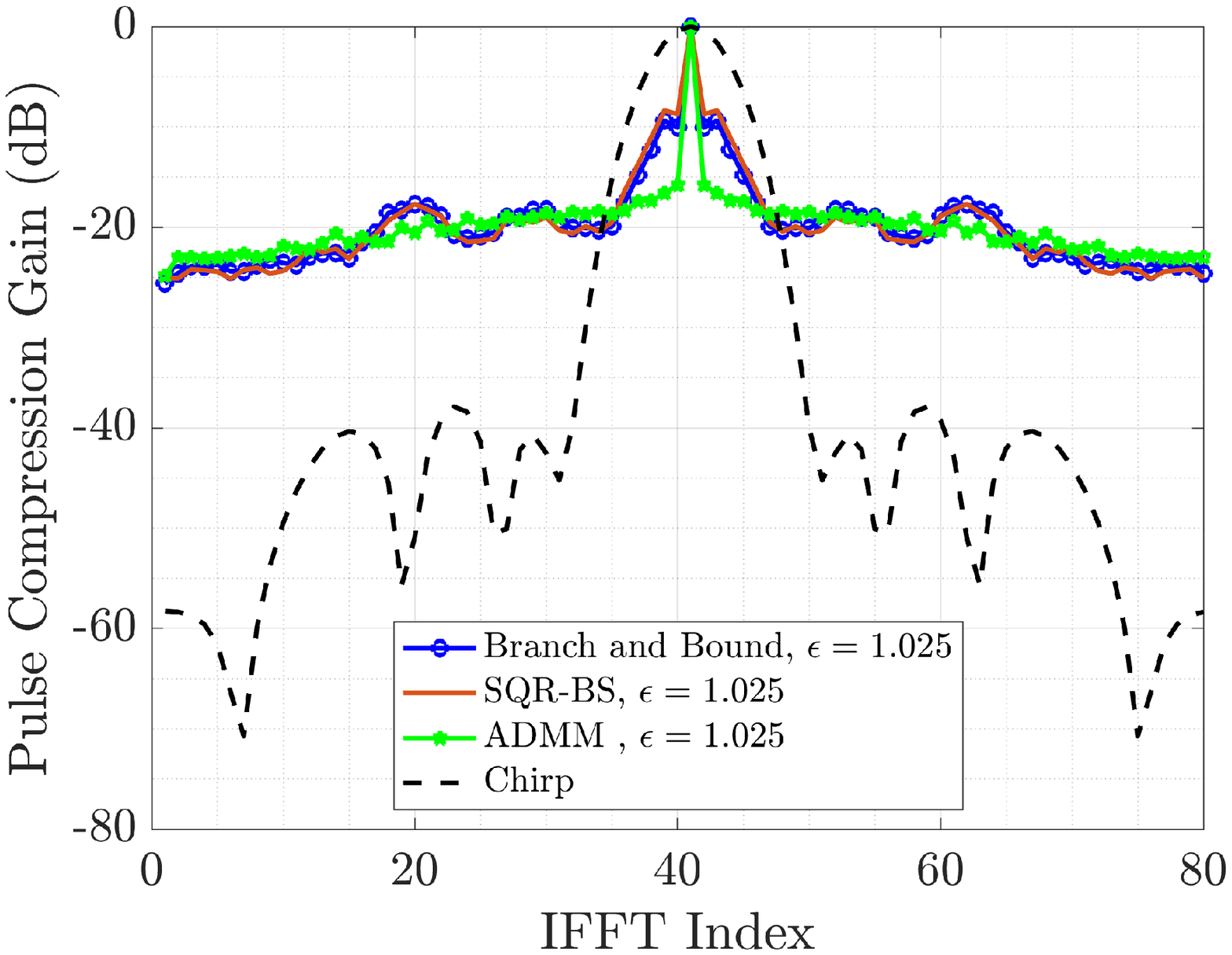}%
\label{fig:PulseGainCompression_c}}
\caption{The pulse compression gain associated to the generated waveform through different approaches and different values of $\epsilon$ for $N = 4$ antennas and $K = 2$ communication users. (a) $\epsilon  = 0.05$ (b) $\epsilon  = 0.44$ (c) $\epsilon  = 1.025$.}
\label{fig:PulseGainCompression}
\end{figure*}


\begin{figure}[t]
	\centering
	\includegraphics[width=1\linewidth]{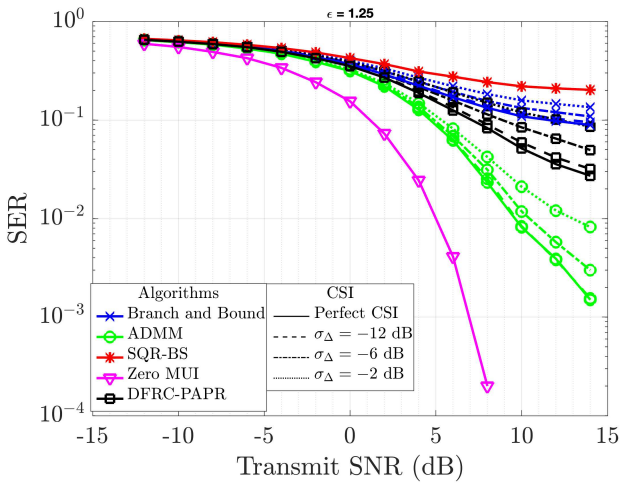}
	\caption{\textcolor{black}{Symbol error rate performance at $\epsilon = 1.25$, $\eta = 1$ for $N = 5$ antennas and $K = 2$ communication users in the presence of imperfect \ac{CSI}. The constellation utilized is QPSK.}}
	\label{fig:SER}
\end{figure}

\begin{figure*}[!t]
\centering
\subfloat[]{\includegraphics[width=3.5in]{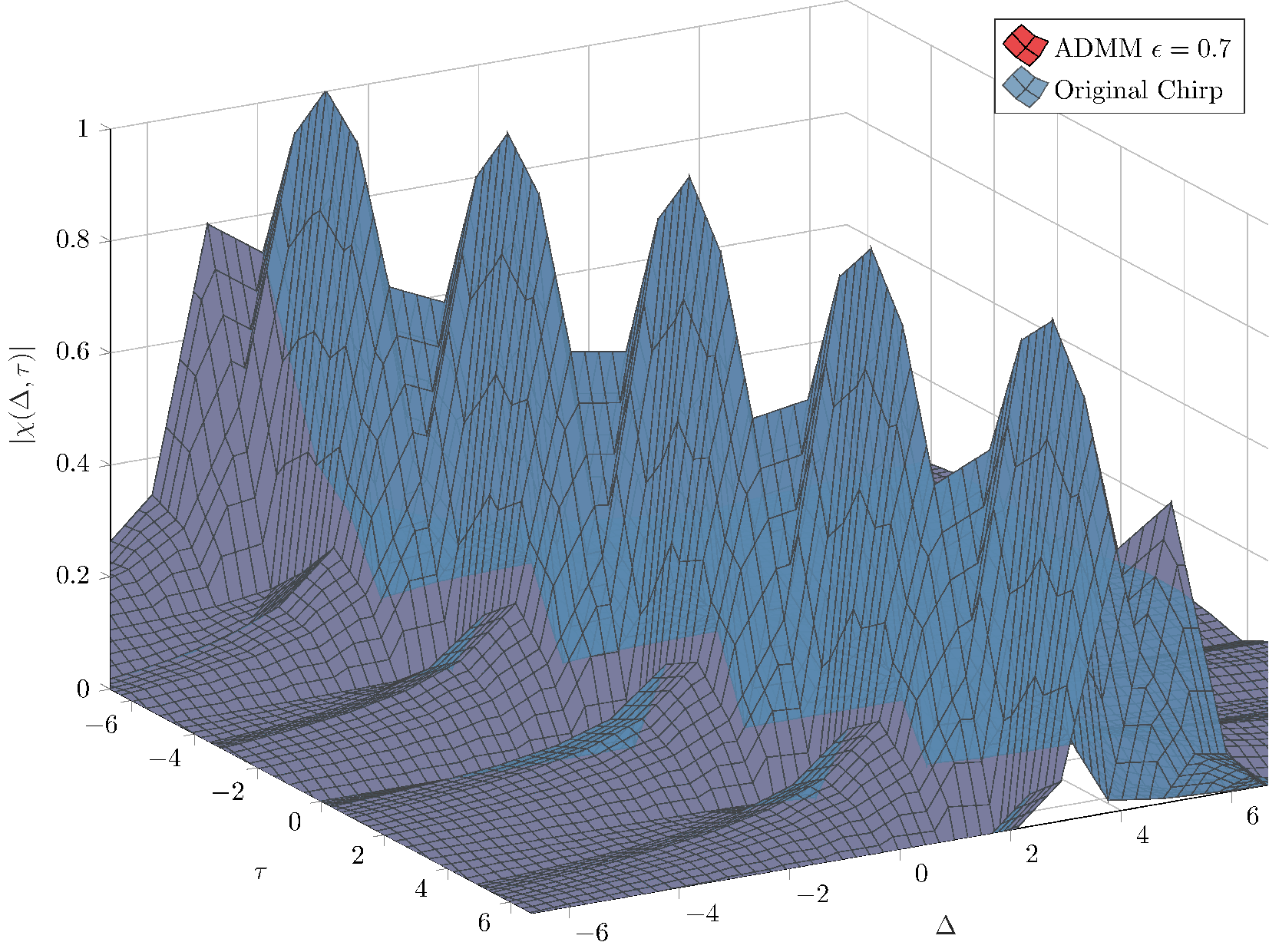}%
\label{fig:AF_a}}
\hfil
\subfloat[]{\includegraphics[width=3.5in]{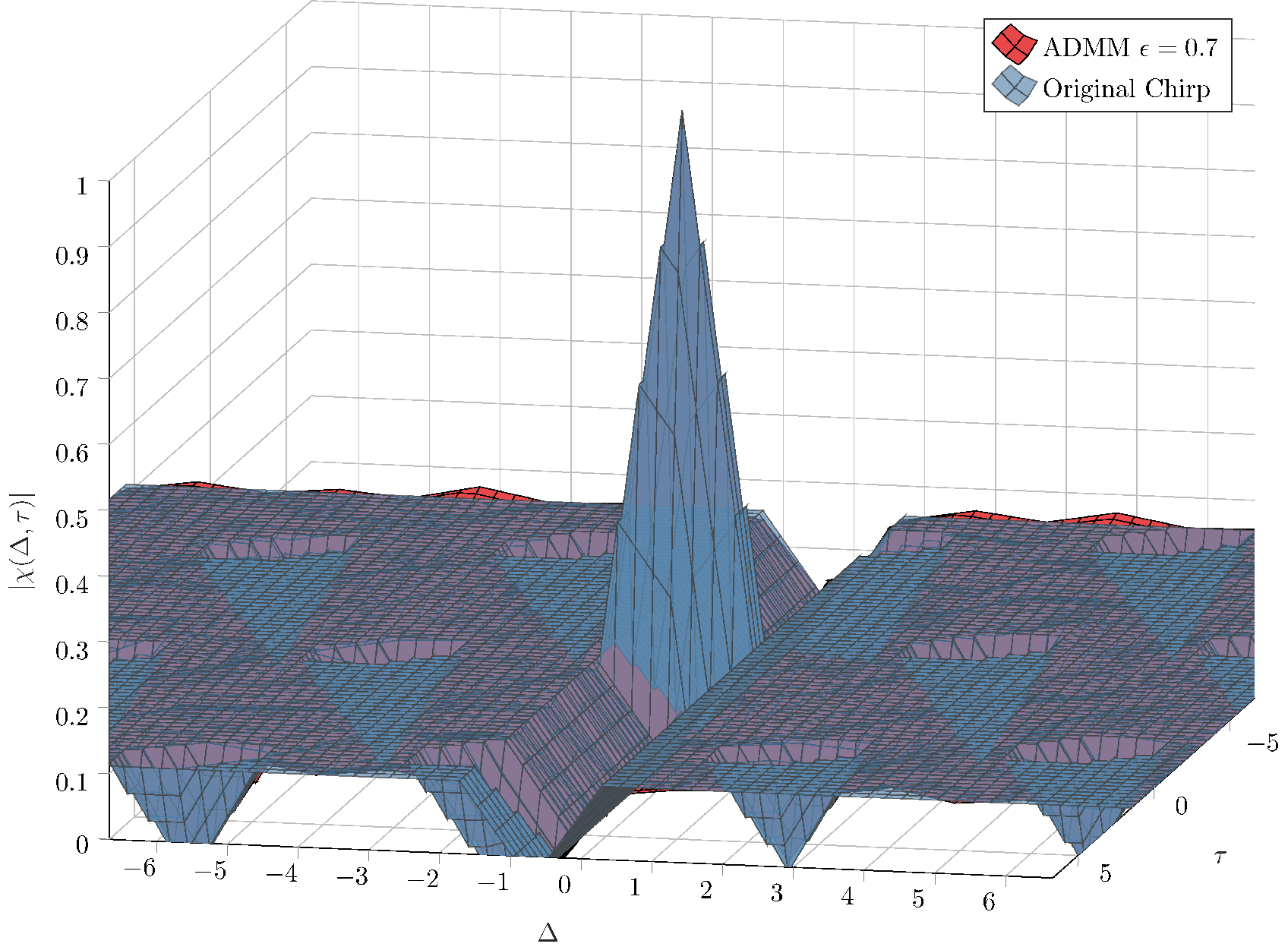}%
\label{fig:AF_b}}
\caption{\textcolor{black}{The ambiguity function $\chi(\Delta,\tau)$ for different chirp types generated comparing the original chirp to the one generated by the \ac{ADMM}-based \ac{DFRC} waveform design approach, when the former is used as a reference signal at $\epsilon = 0.7$. (a) Orthogonal \ac{LFM} chirp; (b)$m$-sequence chirp convolved}}
\label{fig:AF}
\end{figure*}

\section{Simulation Results}
\label{sec:sim}

In this section, a number of simulation results are conducted to illustrate the performance and trade-offs achieved with the proposed DFRC-waveform design. \textcolor{black}{\input{actions01/LandWaveform.tex}}
\vspace{-1cm}
\subsection{Constellations generated with varying $\epsilon$}
\label{subsec:sim-constellations}
We first study the impact of $\epsilon$ on the produced constellation to be used for transmission. Even more, we study its effect on different constellations, as depicted in Fig. \ref{fig:3DConstellations}. We fix $N = 4$ antennas and $\rho = 0.1$ and $\eta = 9$. The number of communication users is set to $K = 2$ and the \ac{ADMM}-based waveform design iterates maximally $M_{{iter}} = 10^3$ iterations. We can observe that a high $\epsilon$ tunes the waveform to be dedicated towards communications. On the other hand, reducing $\epsilon$ introduces distortion onto the transmit constellation, as a tradeoff towards a radar waveform. For example, distortion appears to be present on a QPSK constellation as soon as $\epsilon < 1.6$, compared to $\epsilon < 1.5$ for $16-$QAM. Furthermore, note that for $\epsilon = 2$, the constellation coincides with the output of a zero-forcing equalizer.
\vspace{-0.25cm}
\subsection{Cost convergence}
\label{subsec:simulation-cost-convergence}
In Fig. \ref{fig:CostvsIter}, we aim at studying the convergence of  \textcolor{black}{\textbf{Algorithm \ref{alg:cap}}} over $10^3$ Monte-Carlo trials per $\eta$ value. \textcolor{black}{\input{actions01/costisnowaveraged.tex}}\input{actions01/sim-cost-convergence}.


\subsection{\ac{PAPR} behaviour}
\label{subsec:sim-papr-behaviour}
The \textcolor{black}{\ac{CCDF}} of the \ac{PAPR} for different values of $\eta$ and $\rho$ is shown in Fig. \ref{fig:CCDF}. The \ac{CCDF} is defined as $\Pr(\PAPR > \gamma)$. It is worth noting that increasing $\rho$ results in a steeper cutoff of the \ac{PAPR} \ac{CCDF} most notable for larger values of $\eta$. For instance, \textcolor{black}{focusing on $256-$QAM}, if we set a \ac{CCDF} probability of $10^{-2}$ and a target \ac{PAPR} of $\eta = 0\dB$, we see that for $\rho = 0.1$, the required $\gamma$ is $\gamma = 4.19\dB$, whereas for $\rho = 1$, the $\gamma$ is decreased by $2\dB$. However, tolerating a higher \ac{PAPR}, this gap is reduced. Indeed, for $\eta = 4.8\dB$, we can see that for $\rho = 1$, the requirement is already satisfied at a \ac{CCDF} probability of $10^{-2}$, as opposed to when $\rho = 0.1$, we see that $\gamma = 6.11\dB$, which reflects a gap of about $1.31\dB$. Increasing $\eta = 7\dB$, we see that the gap in $\gamma$ is further reduced to $0.4\dB$ between $\rho = 1$ and $\rho = 0.1$. Finally, this gap becomes negligible at $\eta = 8.5\dB$. \textcolor{black}{\input{actions01/sim-papr-ccdf-behaviour-end.tex}}



\textcolor{black}{\input{actions01/sim-papr-behaviour.tex}}

\subsection{Communication-radar trade-off}
\label{subsec:CommRadarTradeOff}
In Fig. \ref{fig:SumRate}, we aim at studying the trade-off between the sum-rate for communications and the waveform similarity for radar. \textcolor{black}{For each channel realization, and after obtaining the waveform, we compute the lower-bound of the achievable rate of transmission of the $k^{th}$ user according to $R_k = \log_2( 1 + \SINR_k )$ \cite{mohammed2013per}, where 
\begin{equation}
	\SINR_k
	=
	\frac{\mathbb{E}(\vert \pmb{S}_{k,\ell} \vert^2)}{\mathbb{E} \big( \Vert \pmb{\MUI}_{k,\ell} \Vert_F^2 \big) + \sigma_c^2},
\end{equation}
where the expectation is taken over the time index $\ell$. Then, the average achievable sum rate is computed as $ \frac{1}{K}\sum\limits_{k=1}^K R_k$.} Some benchmarks are employed such as the successive \ac{QCQP} refinement (SQR) binary search (SQR-BS) algorithm proposed in \cite{aldayel2016successive}, the \ac{BnB} method \cite{liu2018toward}, \textcolor{black}{the low PAPR-DFRC method in \cite{hu2022low}}, and the \ac{AWGN} capacity. Fig. \ref{fig:SumRate} depicts the average achievable sum-rate as a function of $\epsilon$ for $\SNR = 10\dB$, $N = 4$, and $K = 2$. For $\epsilon < 1.6$, the performance of the proposed \ac{ADMM} waveform design for $\eta = 1$ outperforms that of the QCQP convex bound\footnote{\textcolor{black}{\input{actions01/qcqp.tex}}} and for $\epsilon < 1.8$ the \ac{ADMM} method outperforms the \ac{BnB} method. Also, all methods outperform the SQR-BS method, which coincides with the findings in \cite{liu2018toward}. Raising the $\eta$ to $1.25$ we see that \ac{ADMM} outperforms \ac{BnB} for any $\epsilon$. Furthermore, we see that the \ac{ADMM} attains the \ac{AWGN} capacity performance at $\epsilon > 1.42$ for $\eta = 3$.\textcolor{black}{\input{actions01/insightRef30_Tradeoff.tex}}The gain of the proposed \ac{ADMM} based design can be explained by the added flexibility of the ability of tuning the \ac{PAPR}, which directly impacts the average achievable sum rate. 


\subsection{Radar Pulse Compression Gain}
\label{subsec:sim-radar-pulse-compression-gain}
In Fig. \ref{fig:PulseGainCompression}, we aim at studying the radar pulse compression for different values of $\epsilon$ and for different methods. For simplicity, we study the waveform transmitted by only one antenna, as the main focus is on the temporal aspect. The classical FFT-IFFT pulse compression method \cite{richards2010principles} with a Taylor window to reduce the power of sidelobes. For small $\epsilon$ (ex. $\epsilon = 0.05$ as in Fig. \ref{fig:PulseGainCompression_a} ), we see that the pulse compression of the proposed \ac{ADMM} method better approximates the pulse compression of the original chirp, especially on the sidelobes. For example, at the $19^{th}$ IFFT index, there is a $10$dB difference between the pulse compression of the proposed method and the original chirp, as opposed to $17$dB compared to \ac{BnB} or SQR-BS. As $\epsilon$ is increased, we notice that the proposed \ac{ADMM} tends to fit the original chirp in the mainlobe, rather than the sidelobes. Also, note that for increasing $\epsilon$, the sidelobes of the \ac{ADMM} appear to be decay linearly on a log-scale. \textcolor{black}{\input{actions01/differences-in-pulse-gain-compression.tex}}

\subsection{Symbol Error Rate}
\label{subsec:simulation-SER}
\textcolor{black}{\input{actions01/SER_imperfect_CSI.tex}}

\subsection{Ambiguity function for different chirps}
\label{subsec:sim-ambiguity}
In Fig. \ref{fig:AF}, we study different ambiguity functions when different chirps are embedded within the \ac{ADMM}-based \ac{DFRC} waveform design approach. \textcolor{black}{In Fig. \ref{fig:AF_a}, we plot the ambiguity function of the orthogonal \ac{LFM} waveform.
We also plot the ambiguity function generated by the \ac{ADMM}-based \ac{DFRC} waveform design method at $\epsilon = 0.7$.} It is clearly observed that the ambiguity function given by the \ac{ADMM} design well approximates that of the original orthogonal \ac{LFM} waveform chirp. The same could be said when this chirp waveform is replaced by an $m-$sequence, as shown in Fig. \ref{fig:AF_b}. It is worth noting that the $m-$sequence is utilized for ultra-wide band radars \cite{smeenk2022localization} due to its good ambiguity function properties \cite{berggren2022joint}.

\section{Conclusions}
\label{sec:conclusion}
In this paper, we have proposed an \ac{ADMM}-based \ac{DFRC} waveform design method. The method aims at minimizing the multi-user interference caused by multi-user operation sharing the same spectrum for communications, under radar chirp waveform similarities and peak-to-average-power constraints. The method enjoys flexibility, in a sense that the \ac{PAPR} could be tuned to a desired level, which is of high interest within communication and radar systems, where power amplifiers are part of the transmit chain.  Our analysis reveals that the proposed \ac{ADMM} design is guaranteed to converge to a stable solution. Furthermore, simulation results unveil the superiority of the proposed \ac{ADMM}-based \ac{DFRC} waveform design method, as compared to state-of-the-art radar-communication waveform designs.

Future research will be oriented towards \ac{ISAC} with estimation aspects within the optimization framework. Furthermore, generalizations towards the multi-\ac{DFRC} \textcolor{black}{\ac{BS}} case is also another direction. A possible direction will also be to leverage deep learning techniques for waveform design with various features including but not limited to \ac{PAPR} and radar properties.

\appendices
\small
\label{sec:appendix}
\section{\textcolor{black}{Minimizer of $g^{(m+1)}(\pmb{x})$}}
\label{lemma1-proof}
Note that at the $(m+1)^{th}$ iteration, \ac{ADMM} targets the gradient expressed in equation \eqref{eq:gradient-at-x}, that is also expressed as \textcolor{black}{$\nabla \mathcal{L}_\rho(\bar{\pmb{x}},\pmb{\lambda}^{(m)},\pmb{\mu}^{(m)})$}
\textcolor{black}{
\begin{equation}
\label{eq:gradient-643}
	\nabla_{\pmb{x}} f(\pmb{x}) \big\vert_{\pmb{x}=\pmb{\bar{x}}^{(m+1)}} + \pmb{A}^T \pmb{\lambda}^{(m)} + \rho \pmb{A}^T (\pmb{A} \bar{\pmb{x}}^{(m+1)} - \pmb{\mu}^{(m)} - \pmb{c}) = \pmb{0}.
\end{equation}
}
Using the update equation expressions appearing in equations \eqref{eq:u-update}, \eqref{eq:v-update} and \eqref{eq:w-update}, namely, 
\textcolor{black}{
\begin{equation}
\label{eq:update-the-duals}
\pmb{\lambda}^{(m+1)} = \pmb{\lambda}^{(m)} + \rho (\pmb{A} \bar{\pmb{x}}^{(m+1)} -\pmb{\mu}^{(m+1)} - \pmb{c}).
\end{equation}
}
The gradient expression in equation \eqref{eq:gradient-643} could be reformulated as
\textcolor{black}{ 
\begin{equation}
\label{eq:gradient-743}
\begin{split}
	\nabla_{\pmb{x}} f(\pmb{x}) \big\vert_{\pmb{x}=
	\pmb{\bar{x}}^{(m+1)}}
	& + \pmb{A}^T \Big(\pmb{\lambda}^{(m+1)} - \rho (\pmb{A} \bar{\pmb{x}}^{(m+1)} - \pmb{\mu}^{(m+1)} - \pmb{c}) \Big)  \\
	& + \rho \pmb{A}^T (\pmb{A} \bar{\pmb{x}}^{(m+1)} - \pmb{\mu}^{(m)} - \pmb{c}) = \pmb{0}.
\end{split}
\end{equation}
}
By eliminating common terms we have that, 
\begin{equation}
\label{eq:gradient-alternative-expression}
	\nabla_{\pmb{x}} f(\pmb{x}) \big\vert_{\pmb{x}=
	\pmb{\bar{x}}^{(m+1)}}
	 + \pmb{A}^T \Big(\pmb{\lambda}^{(m+1)} + \rho  (\pmb{\mu}^{(m+1)} - \pmb{\mu}^{(m)}) \Big) = \pmb{0}.
\end{equation}
We can interpret equation \eqref{eq:gradient-alternative-expression} as the gradient of $g(\bar{\pmb{x}})$ where 
\textcolor{black}{
\begin{equation}
	g^{(m+1)}(\pmb{x}) = f(\pmb{x}) 
	 + \pmb{x}^T \pmb{A}^T \Big(\pmb{\lambda}^{(m+1)} + \rho  (\pmb{\mu}^{(m+1)} - \pmb{\mu}^{(m)}) \Big),
\end{equation}
}
with $\bar{\pmb{x}}^{(m+1)}$ being its minimizer.
\vspace{-0.25cm}
\section{Minimizer of $h^{(m+1)}(\pmb{\mu})$}
\label{lemma2-proof}
At the $(m+1)^{th}$ iteration, deriving equation \eqref{eq:another-expression-of-Lrho} with respect to $\pmb{\mu}$, and evaluating $\pmb{x}$ at $\pmb{\bar{x}}^{(m+1)}$ and $\pmb{\lambda}$ at $\pmb{\lambda}^{(m)}$, \textcolor{black}{i.e.}
\begin{equation}
\label{eq:gradient-for-h}
\nabla_{\pmb{\mu}} 
 \mathcal{L}_\rho
  \big\vert_{\pmb{\mu}=
	\pmb{\mu}^{(m+1)}}
=
	-\pmb{\lambda}^{(m)} - \rho (\pmb{A} \bar{\pmb{x}}^{(m+1)} - \pmb{\mu}^{(m+1)}- \pmb{c})
=
\pmb{0}.
\end{equation}
Now using equation \eqref{eq:update-the-duals} in \eqref{eq:gradient-for-h} we get
\begin{equation}
\begin{split}
-\Big(\pmb{\lambda}^{(m+1)} &- \rho (\pmb{A} \bar{\pmb{x}}^{(m+1)} - \pmb{\mu}^{(m+1)} - \pmb{c}) \Big) \\ &- \rho (\pmb{A} \bar{\pmb{x}}^{(m+1)} - \pmb{\mu}^{(m+1)}- \pmb{c})
=
\pmb{0}.
\end{split}
\end{equation}
Through straightforward manipulations, the following function  
\begin{equation}
\label{eq:last-eq-in-app-B}
	h^{(m+1)}(\pmb{\mu}) =- \pmb{\mu}^T  \pmb{\lambda}^{(m+1)} ,
\end{equation}
admits \textcolor{black}{ $\pmb{\mu}^{(m+1)}$ } as a minimizer. Note that function $h$ is super-scripted by $(m+1)$ to emphasize its dependency on $\pmb{\lambda}^{(m+1)}$.
\section{Proof of lemma 3}
\label{lemma3-proof}
We have the following series of equations that hold true for any $\pmb{\lambda}$, namely
\begin{equation}
	\begin{split}
		&(\pmb{\lambda}^{(m+1)} - \pmb{\lambda})^T \pmb{e}^{(m+1)} \\ 
		= &(\pmb{\lambda}^{(m)} + \rho \pmb{e}^{(m+1)}  - \pmb{\lambda})^T \pmb{e}^{(m+1)} \\
		= &(\pmb{\lambda}^{(m)} - \pmb{\lambda}^*)^T \pmb{e}^{(m+1)} + \rho \Vert \pmb{e}^{(m+1)} \Vert^2 \\
		= &\frac{1}{\rho} (\pmb{\lambda}^{(m)} - \pmb{\lambda})^T (\pmb{\lambda}^{(m+1)} - \pmb{\lambda}^{(m)})+ \rho \Vert \pmb{e}^{(m+1)} \Vert^2 \\
		= &\frac{1}{\rho} (\pmb{\lambda}^{(m)} - \pmb{\lambda})^T (\pmb{\lambda}^{(m+1)} - \pmb{\lambda}^{(m)})+ \frac{\rho}{2} \Vert \pmb{e}^{(m+1)} \Vert^2 + \frac{\rho}{2} \Vert \pmb{e}^{(m+1)} \Vert^2 \\
		= &\frac{1}{\rho} (\pmb{\lambda}^{(m)} - \pmb{\lambda})^T (\pmb{\lambda}^{(m+1)} - \pmb{\lambda}^{(m)}) \\ 
		+ &  \frac{1}{2\rho} \Vert \pmb{\lambda}^{(m+1)} - \pmb{\lambda}^{(m)} \Vert^2 
		+ \frac{\rho}{2} \Vert \pmb{e}^{(m+1)} \Vert^2 \\
		= & \frac{1}{2\rho}(\Vert \pmb{\lambda}^{(m+1)} - \pmb{\lambda} \Vert^2  - \Vert \pmb{\lambda}^{(m)} - \pmb{\lambda} \Vert^2) + \frac{\rho}{2} \Vert \pmb{e}^{(m+1)} \Vert^2 .
	\end{split}
\end{equation}
\vspace{-0.7cm}
\section{Proof of Consequence 1}
\label{consequence1-proof}
The following is a direct application of \textbf{Lemma 2} on functions $h^{(m)}(\pmb{\mu})$ and $h^{(m+1)}(\pmb{\mu})$ as follows $-h^{(m)}(\pmb{\mu}^{(m)}) \geq - h^{(m)}(\pmb{\mu}^{(m+1)})$ and $h^{(m)}(\pmb{\mu}^{(m+1)}) \leq h^{(m+1)}(\pmb{\mu}^{(m)})$ along with equation \eqref{eq:update-the-duals}, we get
\begin{equation}
	 -(\pmb{\mu}^{(m+1)})^T \pmb{e}^{(m+1)} 
	 \leq
	 -(\pmb{\mu}^{(m)})^T  \pmb{e}^{(m+1)} .
\end{equation}
\vspace{-0.7cm}
\section{Proof of Theorem 1}
\label{theorem1-proof}

Using \textbf{Lemma 1} and \textbf{Lemma 2}, we can write
\begin{equation}
	g(\pmb{\bar{x}}^{(m+1)}) + h(\pmb{\mu}^{(m+1)})  \leq g(\pmb{\bar{x}}^*) + h(\pmb{\mu}^*).
\end{equation}
After straightforward manipulations, we express the above as
\begin{equation}
\label{eq:lemma1+lemma2}
\begin{split}
	& f(\pmb{\bar{x}}^{(m+1)}) - f(\pmb{\bar{x}}^{*})
	 \\  \leq
	 & -(\pmb{\lambda}^{(m+1)})^T \pmb{e}^{(m+1)}  \\ &
	+\rho(\pmb{\mu}^{(m+1)} - \pmb{\mu}^{(m)})^T (-\pmb{e}^{(m+1)} - (\pmb{\mu}^{(m+1)} - \pmb{\mu}^*)),
\end{split}
\end{equation}
where we have used the fact that $\pmb{e}^* = 0$. Now, using the definition of saddle point, i.e. $\mathcal{L}_0(\bar{\pmb{x}}^*,\pmb{\lambda}^*,\pmb{\mu}^*) \leq \mathcal{L}_0(\bar{\pmb{x}}^{(m+1)},\pmb{\lambda}^*,\pmb{\mu}^{(m+1)})$, we get that $ - (\lambda^*)^T\pmb{e}^{(m+1)} \leq f(\pmb{\bar{x}}^{(m+1)}) -f(\pmb{\bar{x}}^{*})$. Using this as a lower bound on equation \eqref{eq:lemma1+lemma2}, we get
\begin{equation}
\label{eq:lemma1+lemma2+lowerbounded}
\begin{split}
	- &(\pmb{\lambda}^*)^T\pmb{e}^{(m+1)}
	  \leq
	-(\pmb{\lambda}^{(m+1)})^T \pmb{e}^{(m+1)} \\ &
	+\rho(\pmb{\mu}^{(m+1)} - \pmb{\mu}^{(m)})^T (-\pmb{e}^{(m+1)} -(\pmb{\mu}^{(m+1)} - \pmb{\mu}^*)).
\end{split}
\end{equation}
Now, applying Lemma 3 at the saddle point $\pmb{\lambda}^*$, we get the following bound
\begin{equation}
\label{eq:lemma1+lemma2+lowerbounded+S}
\begin{split}
& \frac{1}{2\rho}(\Vert \pmb{\lambda}^{(m+1)} - \pmb{\lambda}^{*} \Vert^2  - \Vert \pmb{\lambda}^{(m)} - \pmb{\lambda}^{*} \Vert^2) + \frac{\rho}{2} \Vert \pmb{e}^{(m+1)} \Vert^2 \\ & 
\leq
\rho(\pmb{\mu}^{(m+1)} - \pmb{\mu}^{(m)})^T (-\pmb{e}^{(m+1)} - (\pmb{\mu}^{(m+1)} - \pmb{\mu}^*)) \\ &
=-\rho(\pmb{\mu}^{(m+1)} - \pmb{\mu}^{(m)})^T \pmb{e}^{(m+1)} + T_m.
\end{split}
\end{equation}
We can further manipulate $T_m$ by adding and subtracting the term $\frac{\rho}{2} \Vert \pmb{\mu}^{(m+1)} - \pmb{\mu}^{(m)} \Vert^2$. This will enable us to express it as a sum of norms,
\begin{equation}
\label{eq:S-manipulation}
	\begin{split}
		T_m &= - \rho (\pmb{\mu}^{(m+1)} - \pmb{\mu}^{(m)})^T(\pmb{\mu}^{(m+1)} - \pmb{\mu}^*) \\
		&-\frac{\rho}{2} \Vert \pmb{\mu}^{(m+1)} - \pmb{\mu}^{(m)} \Vert^2 +\frac{\rho}{2} \Vert \pmb{\mu}^{(m+1)} - \pmb{\mu}^{(m)} \Vert^2 \\
		&= -\frac{\rho}{2}( \Vert \pmb{\mu}^{(m+1)} - \pmb{\mu}^{*} \Vert^2 -\Vert \pmb{\mu}^{(m)} - \pmb{\mu}^{*} \Vert^2  ) \\ &
		-\frac{\rho}{2} \Vert \pmb{\mu}^{(m+1)} - \pmb{\mu}^{(m)} \Vert^2.
	\end{split}
\end{equation}
Using equation \eqref{eq:S-manipulation} in equation \eqref{eq:lemma1+lemma2+lowerbounded+S} we can now say that
\begin{equation}
\label{apply-lemma-to-bound}
	\begin{split}
		 U_{m+1} - U_m & \leq -\rho(\pmb{\mu}^{(m+1)} - \pmb{\mu}^{(m)})^T \pmb{e}^{(m+1)} \\ & - \frac{\rho}{2} \Vert \pmb{e}^{(m+1)} \Vert^2 -\frac{\rho}{2} \Vert \pmb{\mu}^{(m+1)} - \pmb{\mu}^{(m)} \Vert^2,
	\end{split}
\end{equation}
where $U_m = \frac{1}{2\rho}\Vert \pmb{\lambda}^{(m)} - \pmb{\lambda}^{*} \Vert^2 + \frac{\rho}{2}\Vert \pmb{\mu}^{(m)} - \pmb{\mu}^{*} \Vert^2$.
By applying \textbf{Consequence 1}, equation \eqref{apply-lemma-to-bound} directly implies that $U_m$ is a non-increasing sequence of positive numbers. Now, denoting the sequence $a_m = \frac{\rho}{2} \Vert \pmb{e}^{(m)} \Vert^2 + \frac{\rho}{2} \Vert \pmb{\mu}^{(m+1)} - \pmb{\mu}^{(m)} \Vert^2$ and summing equation \eqref{apply-lemma-to-bound}, we get a partial sum as $S_{M_{{iter}}}=\sum\limits_{m=0}^{M_{{iter}}} a_m < U_0 - U_{M_{{iter}}} < U_0 < \infty$. Now since $a_m \geq 0$, then $S_{M_{{iter}}}$ is a positive series for all $M_{{iter}}$. Taking ${M_{{iter}}} \rightarrow \infty$, this means that the infinite positive series $S_\infty$ is bounded above by $U_0$, hence $S_\infty$ converges. This finally means that $\lim\limits_{m \rightarrow \infty} a_m = 0$. But since $a_m$ is composed of two independent norms, then each norm tends to zero (in the Frobenius sense) with increasing number of iterations. Combining the convergence of both norms finalizes the proof. 
%

\vspace{-0.1cm}
\section*{Acknowledgment}
\textcolor{black}{\input{actions01/acknowledge.tex}}
The authors acknowledge that simulations were done on NYU Abu Dhabi's HPC Jubail Cluster. 

\vspace{-0.1cm}
\bibliographystyle{IEEEtran}
\bibliography{refs.bib}

\end{document}

%% file: acronyms.tex
\begin{acronym}
	\acro{OFDM}{orthogonal frequency-division multiplexing}
	\acro{DL}{downlink}
    \acro{HPA}{high power amplifier}
    \acro{IBO}{input power back-off}
    \acro{MIMO}{multiple-input, multiple-output}
    \acro{PAPR}{peak-to-average power ratio}
    \acro{AWGN}{additive white Gaussian noise}
    \acro{KPI}{key performance indicator}
    \acro{KPIs}{key performance indicators}  
    \acro{D2D}{device-to-device}
    \acro{ISAC}{integrated sensing and communications}
    \acro{DFRC}{dual-functional radar and communication}
    \acro{AoA}{angle of arrival}
    \acro{ToA}{time of arrival}
    \acro{EVM}{error vector magnitude}
    \acro{CPI}{coherent processing interval}
    \acro{eMBB}{enhanced mobile broadband}
    \acro{URLLC}{ultra-reliable low latency communications}
    \acro{mMTC}{massive machine type communications}
	\acro{QCQP}{quadratic constrained quadratic programming}
	\acro{BnB}{branch and bound}
	\acro{SER}{symbol error rate}
	\acro{LFM}{linear frequency modulation}
	\acro{ADMM}{alternating direction method of multipliers}
	\acro{CCDF}{complementary cumulative distribution function}
	\acro{MISO}{multiple-input, single-output}
	\acro{CSI}{channel state information}
	\acro{LDPC}{low-density parity-check}
	\acro{BCC}{binary convolutional coding}
	\acro{IEEE}{Institute of Electrical and Electronics Engineers}
	\acro{ULA}{uniform linear antenna}
	\acro{B5G}{beyond 5G}
	\acro{MOOP}{multi-objective optimization problem}
	\acro{RF}{radio frequency}
	\acro{AM/AM}{amplitude modulation/amplitude modulation}
	\acro{AM/PM}{amplitude modulation/phase modulation}
	\acro{BS}{base station}
	\acro{MM}{majorization-minimization}
	\acro{LNCA}{$\ell$-norm cyclic algorithm}
	\acro{OCDM}{orthogonal chirp-division multiplexing}
	\acro{TR}{tone reservation}
	\acro{COCS}{consecutive ordered cyclic shifts}
	\acro{LS}{least squares}
	\acro{CVE}{coefficient of variation of envelopes}
	\acro{BSUM}{block successive upper-bound minimization}
	\acro{ICE}{iterative convex enhancement}
	\acro{SOA}{sequential optimization algorithm}
	\acro{BCD}{block coordinate descent}
	\acro{SINR}{signal to interference plus noise ratio}
	\acro{MICF}{modified iterative clipping and filtering}
	\acro{PSL}{peak side-lobe level}
	\acro{SDR}{semi-definite relaxation}
	\acro{KL}{Kullback-Leibler}
	\acro{ADSRP}{alternating direction sequential relaxation programming}
\end{acronym}

%% file: actions01/B5G6G.tex
One important enabling technology for 6G networks is \ac{DFRC}, which not only benefits from shared spectrum and power efficiency, but also from hardware efficiency through \ac{ISAC} shared waveforms \cite{chafii2022ten}. \ac{DFRC} has emerged as a potential 6G technology, due to its dual nature of incorporating radar and communication capabilities. 
From a communications perspective, high data rates, of around $10$ Gbps for 5G \cite{8985528} and $1$ Tbps for 6G \cite{8412482}, are required for \ac{eMBB} use cases, such as augmented reality and video streaming.
Furthermore, \ac{mMTC} targets the deployment of one million device per km$^2$, and requires low-cost, low-power and low-range devices \cite{chafii2022ten}. 
Moreover, with the aim of one milli-second latency, \ac{URLLC} serves mission-critical applications, for example, remote robotic surgery and autonomous driving.
Meanwhile, radar sensing and localization are new functions in 6G, signaling an important advancement for connected intelligence. In terms of sensing accuracy, 6G puts forward stringent localization requirements at a centimeter scale \cite{9376324}.
Given the increasing number of devices by \ac{mMTC}, the extremely high data rates by \ac{eMBB}, the low latency requirements by \ac{URLLC}, as well as the extreme centimeter accuracy by sensing systems in 6G, a natural concern to raise is whether radar and communication systems should be deployed separately or integrated into one \ac{DFRC} system. An independent deployment approach for sensing and \ac{mMTC}/\ac{eMBB}/\ac{URLLC} communications would provoke spectrum congestion, as well as an increased cost of the entire system.
In spite of the fact that the objectives of sensing and communications tasks are contrasting in nature, a promising solution is the joint integration of sensing and communication, thus leading to a \ac{DFRC} system with the intent of optimizing both objectives. Thanks to the utilization of the same wireless infrastructure, spectrum and radio hardware, \ac{DFRC} can unleash the full potential of 6G systems by concurrently meeting the criteria for radar sensing, as well as use case requirements for eMBB, URLLC, and mMTC.

%% file: actions01/literature.tex
The work in \cite{9725261} designs low-\ac{PAPR} \ac{OFDM} waveforms for RadCom system, where communication and radar bands are separated, and \ac{LNCA}, based on \ac{MM}, is adopted to reduce the waveform’s \ac{PAPR}. On a coding level, a self-disarrange Golay block coding algorithm for DFRC OFDM has been proposed in \cite{8936391}. Meanwhile, a \ac{TR} based on the \ac{COCS} of P4 sequence was used to reduce the \ac{PAPR} for \ac{OFDM} RadCom systems in \cite{8647263}. Moreover, \ac{PAPR} reduction, through an iterative clipping scheme based on gradient-descent, was employed for communication-embedded \ac{OCDM} for radar-communication waveforms \cite{8484372}. In \ac{DFRC} designs, \cite{liu2018toward} proposes a constant modulus design to address for both radar and communication metrics through a \ac{BnB} method. Even though the solution is attractive in terms of \ac{PAPR},  it lacks flexibility in yielding a desired \ac{PAPR}. Moreover, the work in \cite{hu2022low} investigates \ac{MIMO}-OFDM waveforms for \ac{DFRC} systems with low PAPR.  Furthermore, the work in \cite{hu2022low} focuses on spatial beamforming for the radar sub-system. In addition, the problem formulation in \cite{hu2022low} considers a weighted combination of communication and radar metrics, i.e. a weighted combination of the multi-user interference for communications and beampattern design for radar. Furthermore, \cite{7132704} suggests \ac{CVE} as a metric to quantify envelope fluctuations for \ac{OFDM} waveform designs, and an iterative \ac{LS} algorithm to lower the \ac{PAPR}. Moreover, \cite{9399818} focuses on transmit and receive beamforming in \ac{OFDM}-\ac{MIMO} \ac{DFRC} via \ac{KL} divergence metrics, and an \ac{ADSRP} algorithm is proposed. In \ac{MIMO} radar, \cite{8141978} generalizes \cite{8239836} to include \ac{PAPR} constraints for radar-return \ac{SINR} output maximization via the \ac{BSUM} method.  Additionally, \cite{cui2013mimo} uses \ac{SOA} and \cite{cui2016space} uses \ac{BCD} to tackle the same problem. Also, \cite{7736116,cui2013mimo,imani2016sequential} jointly designs the transmit sequences and receive filters under \ac{PAPR} constraints. Furthermore, \cite{7707413} controls \ac{PSL} of \ac{MIMO} radar subject to transmit \ac{PAPR} constraint via chaotic waveforms as  initial sequences. In \ac{MIMO} \ac{OFDM}, the designs in \cite{8496888,7272868,6857420} utilize \ac{MICF} to reduce the \ac{PAPR} of the transmitted \ac{OFDM} pulses. From a communication-only perspective, numerous methods could be classified into two classes: \textit{(i)} distorted, such as companding and clipping, and \textit{(ii)} distortionless techniques, for example \ac{TR}. Also, the work in \cite{wang2010optimized} derives an iterative clipping and filtering in frequency domain via convex optimization techniques. An overview of \ac{PAPR} reduction methods for communications are found in \cite{6476061,1421929} and references therein. In contrast to all the previous methods, our work is the first to consider multi-user interference under radar similarity and a controllable \ac{PAPR}.

%% file: actions01/contribution-complexity.tex
\textbf{Computational Complexity Analysis}. We also provide a computational complexity analysis of the proposed \ac{DFRC} waveform design method and show that the proposed method has lower complexity than the state-of-the-art ones.

%% file: actions01/contribution-CSI.tex
\textbf{Imperfect \ac{CSI} Aware Design}. To cope with channel estimation errors and imperfections, we offer a variant of the waveform design algorithm that is aware of imperfect \ac{CSI} and allude to its robustness, as compared to other state-of-the-art methods.

%% file: actions01/claim-proof.tex
The convergence analysis shows that the proposed method is guaranteed to converge, i.e. we can always construct a suitable \ac{ISAC} waveform with a desired \ac{PAPR} and chirp similarity that is also deemed suitable for communications.

%% file: actions01/organization.tex
The paper is organized as follows: We introduce the communication and radar model in Section \ref{sec:systemmodel}. Section \ref{sec:framework} introduces the \ac{ISAC} \ac{DFRC} waveform design optimization problem. Furthermore, we show the impact of \ac{PAPR} on both sensing and communication performances in Section \ref{sec:impact}. Section \ref{sec:admm} presents the \ac{ADMM} based \ac{DFRC} waveform design solution. The convergence analysis is presented in Section \ref{sec:convergence}, whereas Section \ref{sec:complexity} presents our complexity analysis. Section \ref{sec:imperfect-CSI} extends the proposed \ac{DFRC} waveform design solution to cope with imperfect \ac{CSI}. Section \ref{sec:sim} illustrates our simulation findings, and Section \ref{sec:conclusion} concludes our work.


%% file: actions01/arrayresponse.tex
\begin{equation}
	\label{eq:yr}
	\pmb{Y}_r = \gamma_0 \pmb{a}_{N_R}(\theta_0)\pmb{a}_{N}^T(\theta_0)\pmb{X} +  \sum\limits_{n=1}^C \gamma_n \pmb{a}_{N_R}(\theta_n)\pmb{a}_N^T(\theta_n)\pmb{X} + \pmb{Z}_r,
\end{equation}
where $N_R$ is the number of receiving antennas. The vectors $\pmb{a}_{N}(\theta) \in \mathbb{C}^{N \times 1}$ and $\pmb{a}_{N_R}(\theta) \in \mathbb{C}^{N_R \times 1}$ represent the transmit and receive array steering vectors at angle $\theta$, respectively. For example, if the antenna configuration follows a \ac{ULA} array, then the steering vector can be expressed as
\begin{equation}
	\pmb{a}_N
	(\theta)
	=
	\begin{bmatrix}
		1 & e^{j \frac{2\pi}{\lambda} d \sin(\theta)} & \ldots & e^{j \frac{2\pi}{\lambda} d(N-1) \sin(\theta)}
	\end{bmatrix}^T,
\end{equation}
where $\lambda$ is the wavelength and $d$ is the inter-element spacing between antennas. Due to the mono-static setting, the angle of departure and angle of arrival of the different echo components are the same.

%% file: actions01/do-we-use-theta.tex
In this article, we focus on the design of the waveform $\pmb{X}$ under  similarity constraints relative to a given chirp with desirable auto-correlation properties. Therefore, the radar beamforming design is not the main focus of this paper. 

%% file: actions01/motivation-time-samples.tex
The motivation behind using time samples for DFRC signals is two-fold. From a communication perspective, the transmit signal encodes $PKL$ bits where $2^P$ is the constellation size, thus embedding more bits within the same transmit frame. Additionally, having more time samples within the same transmit signal enables the usage of forward error correction codes, ex. \ac{LDPC} and \ac{BCC} codes that can further enhance the transmission reliability. On the other hand, and from a sensing perspective, the time samples of the DFRC signal allow similarity between the transmit signal and a given chirp waveform with attractive properties such as doppler tolerance and high range resolution.

%% file: actions01/EMUI.tex
According to \cite{mohammed2013per} (cf. equation (30) therein), it has been shown that the MUI directly impacts the achievable sum-rate of the \textcolor{black}{\ac{DL}} users. In particular, a lower bound on the achievable sum-rate of the $k^{th}$ communication user is maximized by minimizing the total MUI energy under fixed transmit constellation energy, which is given by the following expression
\begin{equation}
\label{eq:EMUI}
	E_{\MUI} = \Vert \pmb{\MUI} \Vert_F^2.
\end{equation}
Indeed, $E_{\MUI}$ reflects the amount of energy interfering with symbol detection in an AWGN channel as per equation \eqref{eq:received-signal-2}. The MUI metric has been adopted as a communication metric in different contexts. For example, the work in \cite{liu2020range} minimizes MUI while trying to reduce range sidelobes for \ac{DFRC} systems. Also, \cite{liu2018toward} formulates a weighted optimization problem taking into account MUI minimization. 

%% file: actions01/following-set-of-logic.tex
Using the following set of logic, i.e.
\begin{equation}
	\begin{split}
		E_{\MUI} = \Vert \pmb{H}\pmb{X} - \pmb{S} \Vert_F^2  &= \Vert \pmb{H} ( \pmb{X} - \pmb{H}^\dagger{\pmb{S}}) \Vert_F^2 \\
				 &\leq \Vert \pmb{H} \Vert_F^2   \Vert  \pmb{X} -\pmb{H}^\dagger\pmb{S} \Vert_F^2 \\
				 &= \Vert \pmb{H} \Vert_F^2 \Vert \pmb{x} - \pmb{x}_{\comm} \Vert^2,
	\end{split}
\end{equation}
where $\pmb{x}_{\comm} = \ve(\pmb{H}^H(\pmb{H}\pmb{H}^H)^{-1}\pmb{S})$. Note that we have used $\Vert \pmb{A}\pmb{B} \Vert_F^2 \leq\Vert \pmb{A} \Vert_F^2\Vert \pmb{B} \Vert_F^2 $. Moreover, we aim at minimizing an upper bound part of the MUI that depends on the waveform $\pmb{X}$. Therefore, the rest of the paper deals with the following problem,

%% file: actions01/precodingwaveform.tex
The precoding part is contained within $\pmb{X}$. Indeed, solving the unconstrained version of problem $(\mathcal{P}')$ results in a zero-forcing precoded solution, namely $\hat{\pmb{x}}^{\tt{opt}} = \pmb{x}_{\comm}$ (or equivalently $\hat{\pmb{X}}^{\tt{opt}} = \pmb{H}^H(\pmb{H}\pmb{H}^H)^{-1}\pmb{S}$). In this case, it is clear that the zero-forcing precoded solution would \textit{vanish} for the multi-user \ac{MISO}, thus leading to an \ac{AWGN} channel observed at the communication users. However, the constrained problem in $(\mathcal{P}')$ accounts for radar constraints with limited \ac{PAPR}, which prevents us from explicitly expressing the transmit waveform $\pmb{X}$ as a precoded version of the desired constellation $\pmb{S}$. In the following, we present an algorithm tailored to solve problem $(\mathcal{P}')$ in an efficient manner.

%% file: actions01/PAPR-MUI-Similarity.tex
 In real transmitting \ac{RF} chains, an essential component to maintain a desired transmit power is the \ac{HPA}. \ac{HPA}s are typically the most power-hungry blocks of an \ac{RF} system \cite{777273}. Hypothetically, one desires an ideal \ac{RF} \ac{HPA} to avoid signal clipping, however, linear \ac{HPA}s suffer from power inefficiency, large size, and circuit complexity \cite{1381700}. For non-linear \ac{HPA}s, a naive way to deal with high-\ac{PAPR} waveforms is to tune the \ac{IBO} high enough so that the actual operating point, a.k.a \textit{quiescent point} or \textit{Q-point}, falls within the linear region of the \ac{HPA}. Note that the \textit{quiescent point} is set to match the steady-state DC component of the waveform, i.e. the average power of the waveform. A high \ac{IBO} guarantees operation in the linear zone, but sacrifices low power efficiency and low transmit power. Instead, a sophisticated approach would lower the \ac{IBO} so that the Q-point is as close as possible to the ideal operating point, however, the input waveforms should maintain a low \ac{PAPR} (which is normally done in baseband) to avoid clipping. This leads to the following important question: \textit{What impact does \ac{PAPR} have on the sensing (from a chirp similarity perspective) and communication (from a MUI perspective) performances?} To answer this question, a natural tool  to leverage is \ac{MOOP}. For more information on \ac{MOOP}, the reader is referred to \cite{6924852}.  
 
 In order to jointly optimize for communications and sensing performances, a natural \ac{MOOP} to consider is the following:
\begin{equation}
	\label{eq:MOOP1}
\begin{aligned}
(\mathcal{M}):
\begin{cases}
\min\limits_{\lbrace \pmb{x} \rbrace}&  \begin{bmatrix}
	\phi_1(\pmb{x}) , \phi_2(\pmb{x})
\end{bmatrix} \\
\textrm{s.t.}
 &  \pmb{x} \in \mathbb{C}^{ML},
\end{cases}
\end{aligned}
\end{equation}
where $\phi_1(\pmb{x})=E_{\MUI}$ and $\phi_2(\pmb{x}) = \Vert \pmb{x} - \pmb{x}_0 \Vert^2$. To infer the impact of \ac{PAPR} on the communications and sensing performances, we study the \textit{Pareto boundary} of the \ac{MOOP} in \eqref{eq:MOOP1}. Next, we compute the \textit{Pareto boundary} by passing the optimal waveforms through an \ac{HPA} with an \ac{IBO} that tolerates a maximal \ac{PAPR} of $\PAPR_0$ without clipping. In other words, all waveforms with \ac{PAPR} greater than $\PAPR_0$ are subject to clipping. Another interesting \ac{MOOP} to consider is
\begin{equation}
	\label{eq:MOOP2}
\begin{aligned}
(\mathcal{M}_{\eta}):
\begin{cases}
\min\limits_{\lbrace \pmb{x} \rbrace}&  \begin{bmatrix}
	\phi_1(\pmb{x}) , \phi_2(\pmb{x})
\end{bmatrix} \\
\textrm{s.t.}
 &  \PAPR(\pmb{x}) \leq \eta, \quad  \pmb{x} \in \mathbb{C}^{ML}.
\end{cases}
\end{aligned}
\end{equation}
In equation \eqref{eq:MOOP2}, the \ac{MOOP} is aware of an \ac{HPA} operating at an \ac{IBO} corresponding to a maximal tolerated \ac{PAPR} of $\eta$. Moreover, the \ac{MOOP} in \eqref{eq:MOOP2} is expected to return waveforms with a \ac{PAPR} of at most $\eta$, while optimizing for sensing and communication performances.

In Fig. \ref{fig:Pareto}, we plot the Pareto boundary of the \ac{MOOP}s in equation \eqref{eq:MOOP1} and equation \eqref{eq:MOOP2}. As one can observe, the Pareto boundary obtained by solving the \ac{MOOP} in \eqref{eq:MOOP1} both sensing and communications performance. For example, if we fix the similarity requirement to $1$, the $\MUI$ energy looses approximately $4.6 \dB$ when going from $\PAPR_0 = 8.33\dB$ towards $\PAPR_0 = 1.05\dB$. A similar argument can be made by fixing the $\MUI$ energy. On the other hand, considering the Pareto boundary of $(\mathcal{M}_{\eta})$ in \eqref{eq:MOOP2}, a \textit{significant recovery} of performance can be noticed. Hence, including a $\PAPR$ constraint in the optimization problem can improve the sensing-communication performance, while maintaining a desired $\PAPR$ through $\eta$ for \ac{IBO} considerations of the \ac{HPA}.

%% file: actions01/admm-round-robin-equations.tex
\begin{equation}
\label{eq:update-ADMM-equations-all}
\begin{split}
	\bar{\pmb{x}}^{(m+1)}
	&=
	\arg\min_{\bar{\pmb{x}}}
	\mathcal{L}_{\rho}(\bar{\pmb{x}},
	\begin{bmatrix}
	\pmb{\alpha} ,\pmb{\beta},\pmb{\gamma},\pmb{u},\pmb{v},\pmb{w}
	\end{bmatrix}^{(m)}
	), \\
	\pmb{\alpha}^{(m+1)}
	&=
	\arg\min_{\pmb{\alpha} \in \mathcal{C}_{\alpha}}
	\mathcal{L}_{\rho}(\bar{\pmb{x}}^{(m+1)},\pmb{\alpha},\begin{bmatrix}
	\pmb{\beta},\pmb{\gamma},\pmb{u},\pmb{v},\pmb{w}
	\end{bmatrix}^{(m)}), \\
	\pmb{\beta}^{(m+1)}
	&=
	\arg\min_{\pmb{\beta} \in \mathcal{C}_{\beta}}
	\mathcal{L}_{\rho}(
	\begin{bmatrix}
		\bar{\pmb{x}},\pmb{\alpha}
	\end{bmatrix}^{(m+1)},
	\pmb{\beta},
	\begin{bmatrix}
	\pmb{\gamma},\pmb{u},\pmb{v},\pmb{w}
	\end{bmatrix}^{(m)}), \\
	\pmb{\gamma}^{(m+1)} 
	&=
	\arg\min_{\lbrace \pmb{\gamma}_n \in \mathcal{C}_{\gamma} \rbrace_{n=1}^{NL}}
	\mathcal{L}_{\rho}(\begin{bmatrix}
		\bar{\pmb{x}},\pmb{\alpha},\pmb{\beta}
	\end{bmatrix}^{(m+1)},\pmb{\gamma},\begin{bmatrix}
	\pmb{u},\pmb{v},\pmb{w}
	\end{bmatrix}^{(m)}),
\end{split}
\end{equation}

%% file: actions01/text-below-beta.tex
In order to update $\pmb{\gamma}_n$, all quantities except for $\pmb{\gamma}_n$ are set to their most recent value. In other words, we use $\bar{\pmb{x}}^{(m+1)},\pmb{\alpha}^{(m+1)}$, $\pmb{\beta}^{(m+1)}$, $\pmb{u}^{(m)},\pmb{v}^{(m)}$ and $\pmb{w}_1^{(m)} \ldots \pmb{w}_{NL}^{(m)}$. To this extent, we have the following optimization problem,

%% file: actions01/algorithm1.tex
\begin{algorithm}[H]
\caption{\textcolor{black}{ADMM-based \ac{DFRC} Waveform Design}}\label{alg:cap}
\begin{algorithmic}
\STATE 
\STATE \textcolor{black}{{\textsc{input}:} $\pmb{x}_0, \pmb{H}, \pmb{S}$}
\STATE \textcolor{black}{{\textsc{initialize}:} }
\STATE \hspace{0.5cm} \textcolor{black}{$\pmb{\alpha}^{(0)} \gets \pmb{0}$, $\pmb{\beta}^{(0)} \gets \pmb{0}$, $\pmb{\gamma}_n^{(0)} \gets \pmb{0} \quad \forall n = 1 \ldots NL$.} 
\STATE \hspace{0.5cm} \textcolor{black}{$\pmb{u}^{(0)} \gets \pmb{0}$, $\pmb{v}^{(0)} \gets \pmb{0}$, $\pmb{w}_n^{(0)} \gets \pmb{0} \quad \forall n = 1 \ldots NL$.}
\STATE \hspace{0.5cm} \textcolor{black}{$\pmb{x}_{\comm} \gets \ve(\pmb{H}^H(\pmb{H}\pmb{H}^H)^{-1}\pmb{S})$.}
\STATE \hspace{0.5cm}  \textcolor{black}{$\bar{\pmb{x}}_{0} \gets \begin{bmatrix}
	\Real({\pmb{x}}_{0})^T & \Imag({\pmb{x}}_{0})^T\end{bmatrix}^T$.}
\STATE \hspace{0.5cm} \textcolor{black}{$\bar{\pmb{x}}_{\comm} \gets \begin{bmatrix}
\Real({\pmb{x}}_{\comm})^T & \Imag({\pmb{x}}_{\comm})^T \end{bmatrix}^T$.}
\STATE \hspace{0.5cm} \textcolor{black}{$m \gets 0$}

\STATE\textcolor{black}{ {\textsc{while}} $ m < M_{iter}$}
\STATE \hspace{0.5cm} \textcolor{black}{Update $\bar{\pmb{x}}^{(m+1)}$ using equation \eqref{eq:x-update}}
\STATE \hspace{0.5cm} \textcolor{black}{Update $\pmb{\alpha}^{(m+1)}$ using equation \eqref{eq:alpha-update}}
\STATE \hspace{0.5cm} \textcolor{black}{Update $\pmb{\beta}^{(m+1)}$ using equation \eqref{eq:beta-update} }
\STATE \hspace{0.5cm} \textcolor{black}{Update $\pmb{\gamma}_1^{(m+1)} \ldots \pmb{\gamma}_{NL}^{(m+1)}$ via equation \eqref{eq:gamma-update}}
\STATE \hspace{0.5cm} \textcolor{black}{Update $\pmb{u}^{(m+1)}$ using equation \eqref{eq:u-update}}
\STATE \hspace{0.5cm} \textcolor{black}{Update $\pmb{v}^{(m+1)}$ using equation \eqref{eq:v-update}}
\STATE \hspace{0.5cm} \textcolor{black}{Update $\pmb{w}_1^{(m+1)} \ldots \pmb{w}_{NL}^{(m+1)}$ via  equation \eqref{eq:w-update}}
\STATE \hspace{0.5cm} \textcolor{black}{$m \gets m + 1$}
\STATE\textcolor{black}{\textbf{return}  $\bar{\pmb{x}}^{(M_{{iter}})}$}
\end{algorithmic}
\end{algorithm}

%% file: actions01/complexity.tex
In this section, we analyze the computational complexity of the proposed method described in \textbf{Algorithm 1}. To start with, the initialization phase consists of computing $\pmb{x}_{\comm} = \pmb{H}^H(\pmb{H}\pmb{H}^H)^{-1}\pmb{S}$. This operation costs $\mathcal{O}(2 K^2 N+ K^3 + 2NK^2 + 2NKL)$, where the matrix multiplication $\pmb{H}\pmb{H}^H$ costs $\mathcal{O}(2K^2N)$, the involved inverse costs $\mathcal{O}(K^3)$, and  $\mathcal{O}(2NK^2 + 2NKL)$ come from the left and right matrix multiplications by $\pmb{H}^H$ and $\pmb{S}$, respectively. Furthermore, we observe that updating equations \eqref{eq:x-update}, \eqref{eq:alpha-update}, \eqref{eq:beta-update} and \eqref{eq:gamma-update} require three constant parameters involving divisions, i.e. $\frac{1}{\rho}, \frac{1}{2+3\rho} $ and $\sqrt{\frac{\eta}{NL}}$. In what follows, we assume that these parameters are computed once and stored for usage within the main loop.

In the main loop, updating $\bar{\pmb{x}}^{(m+1)}$ in equation \eqref{eq:x-update} costs $\mathcal{O}(18NL)$ operations. Note that the multiplication $\bar{\pmb{F}}_n\pmb{w}_n^{(m)}$ is a very simple operation due to the fact that $\bar{\pmb{F}}_n$ is a selection matrix, that picks the $n^{th}$ and $(NL+n)^{th}$ entries of $\pmb{w}_n^{(m)}$ and sets all other entries to zero. Therefore, its total cost is $\mathcal{O}(1)$. This means that both summations $\sum\limits_{n=1}^{NL}\bar{\pmb{F}}_n\pmb{w}_n^{(m)}$ and $\sum\limits_{n=1}^{NL}\bar{\pmb{F}}_n\pmb{\gamma}_n^{(m)}$ require $\mathcal{O}(NL)$. 
Moreover, updating $\pmb{\alpha}^{(m+1)}$ by equation \eqref{eq:alpha-update} costs $\mathcal{O}(6NL)$. Furthermore, the worst-case complexity of $\pmb{\beta}^{(m+1)}$ via equation \eqref{eq:beta-update} costs $\mathcal{O}(10NL)$, which happens when $\pmb{\beta}^{(m+1)} \notin  \mathcal{C}_\beta$ as an additional multiplication with $\epsilon$ and division for normalization is required.
Now, we discuss the complexity of updating $\pmb{\gamma}_n$, where $n = 1 \ldots NL$ found in equation \eqref{eq:gamma-update}. Indeed, the worst-case complexity to update $\pmb{\gamma}_n^{(m+1)}$ is $\mathcal{O}(8NL + 2)$, which occurs when $\pmb{\gamma}_n^{(m+1)} \notin  \mathcal{C}_\gamma$. Therefore, updating the entire batch of $\lbrace \pmb{\gamma}_n \rbrace_{n=1}^{NL}$ costs $\mathcal{O}(NL(8NL + 2))$.
Next, it is straightforward to see that the computational complexity for updating $\pmb{u}^{(m+1)}$ according to equation \eqref{eq:u-update} comprises of $\mathcal{O}(6NL)$ flops. Similarly, updating $\pmb{v}^{(m+1)}$ according to equation \eqref{eq:v-update} costs $\mathcal{O}(8NL)$ flops. The last operation in the main loop is to update $\lbrace \pmb{w}_n \rbrace_{n=1}^{NL}$. As previously explained, the operation $\bar{\pmb{F}}_n\bar{\pmb{x}}^{(m+1)}$ is simple due to the selection nature of $\bar{\pmb{F}}_n$. Based on this, we can observe that updating $\pmb{w}_n^{(m+1)}$ costs $\mathcal{O}(4NL+2)$ flops. Finally, updating for all $\pmb{w}_n^{(m+1)}$'s for $n = 1\ldots NL$ costs $\mathcal{O}(NL(4NL+2))$. Adding all costs, we conclude that the overall computational complexity of the proposed \ac{ADMM}-based \ac{DFRC} waveform design algorithm costs
\begin{equation}
			T = 2K^2N + K^3 + 2NK^2 + 2NKL  + M_{iter}(12 N^2 L^2 + 52NL ),
\end{equation}
where $T$ is the total number of flops involved. Using \textit{Big-O analysis}, the worst-case computational complexity of the resulting method scales as $\mathcal{O}( M_{{iter}} N^2 L^2 + K^2 N + K^3 + NK^2 + NKL)$. Therefore, the proposed method is computationally more efficient than the branch and bound method in \cite{liu2018toward}, which requires $\mathcal{O}( 2^{N+1})$, and the successive \ac{QCQP} refinement (SQR) binary search (SQR-BS) in \cite{aldayel2016successive}, which costs $\mathcal{O}( M_{{iter}} N^{3.5} L^{3.5})$.

%% file: actions01/imperfect_CSI.tex
The proposed \ac{ADMM}-based \ac{DFRC} waveform design method is based on the assumption of perfect \ac{CSI}, i.e. the \ac{DFRC} base station has a perfect estimate of the channel matrix $\pmb{H}$. In reality, this is not true as inaccurate estimation and quantization errors are part of transmit and receive paths. Another motivation of imperfect \ac{CSI} is outdated effects. To model imperfect \ac{CSI}, we introduce \ac{CSI} errors integrated within $\pmb{\Delta}$, which is assumed to be deterministic norm-bounded. Therefore, the \ac{CSI} estimate at the \ac{DFRC} is 
\begin{equation}
	\pmb{H}
	=
	\widetilde{\pmb{H}} 
	+
	\pmb{\Delta}.
\end{equation}
The MUI energy in this case is expressed as 
\begin{equation}
	E_{\MUI} = \Vert 
	(\widetilde{\pmb{H}} 
	+
	\pmb{\Delta})\pmb{X} - \pmb{S}
	\Vert_F^2.
\end{equation}
Then, problem $(\mathcal{P})$ can be extended to account for imperfect \ac{CSI} via a norm-bounded perspective, given as follows
\begin{equation}
 \label{eq:problem1-imperfect-csi}
\begin{aligned}
(\mathcal{P}_{\tt{robust}}):
\begin{cases}
\min\limits_{\lbrace \pmb{x} \rbrace}
\max\limits_{ \Vert \pmb{\Delta} \Vert \leq \sigma_{\Delta}}
&  \Vert 
	(\widetilde{\pmb{H}} 
	+
	\pmb{\Delta})\pmb{X} - \pmb{S}
	\Vert_F^2\\
\textrm{s.t.}
 &  \textcolor{black}{\Vert \pmb{x} \Vert^2 = 1,} \\ 
 & \pmb{x}^H \pmb{F}_n \pmb{x} \leq \frac{\eta}{NL} , \quad \forall n \\
 & \pmb{x} \in \mathcal{B}_{\epsilon}(\pmb{x}_0), \\ 
\end{cases}
\end{aligned}
\end{equation} 
where $\sigma_{\Delta}$ sets the norm-bounded, or worst-case, magnitudes on the \ac{CSI} errors $\pmb{\Delta}$. Now, we can upper bound the MUI via 
\begin{equation}
	\begin{split}
\max\limits_{ \Vert \pmb{\Delta} \Vert \leq \sigma_{\Delta}}
   E_{\MUI}
	&\myeqa
	\big(\Vert 
	\widetilde{\pmb{H}}
	\pmb{X}
	-
	\pmb{S}
	\Vert_F
	+
	\sigma_{\Delta}
	\Vert
	\pmb{X}
	\Vert_F
	 \big)^2 \\
	&\myeqb
	\big(\Vert 
	\widetilde{\pmb{H}}
	(
	\pmb{X}
	-
	\widetilde{\pmb{H}}^\dagger
	\pmb{S}
	)
	\Vert_F
	+
	\sigma_{\Delta}
	\Vert
	\pmb{X}
	\Vert_F
	 \big)^2 \\
	 &\myleqc
	\big(\Vert 
	\widetilde{\pmb{H}}
	\Vert_F
	\Vert
	(
	\pmb{X}
	-
	\widetilde{\pmb{H}}^\dagger
	\pmb{S}
	)
	\Vert_F
	+
	\sigma_{\Delta}
	\Vert
	\pmb{X}
	\Vert_F
	 \big)^2 \\
	 &\myeqd 
	 \Vert 
	\widetilde{\pmb{H}}
	\Vert_F^2 
	\big(\Vert\pmb{x} - \widetilde{\pmb{x}}_{\comm} \Vert + 
	\sigma_{\Delta}'\Vert\pmb{x}\Vert \big) ^2,
	\end{split}
\end{equation}
where the proof of (a) follows similar steps as \cite{huang2011robust} (c.f. Lemma 3.1). Furthermore, given that $\widetilde{\pmb{H}}$ is full row rank, step (b) factors $\widetilde{\pmb{H}}$ and step (c) uses $\Vert \pmb{A}\pmb{B}\Vert \leq  \Vert \pmb{A} \Vert \Vert \pmb{B} \Vert$. Finally, step (d) reformulates the MUI in terms of vectorized versions of the involved quantities, i.e. $\pmb{x} = \ve(\pmb{X})$ and $\widetilde{\pmb{x}}_{\comm} = \ve(\widetilde{\pmb{H}}^\dagger
	\pmb{S})$. Also, $\sigma_{\Delta}' = \frac{\sigma_{\Delta}}{\Vert 
	\widetilde{\pmb{H}}
	\Vert_F}$ represents the normalized worst-case magnitudes.   
Therefore, it follows that the robust waveform design problem in equation \eqref{eq:problem1-imperfect-csi} is casted as 
\begin{equation}
 \label{eq:problem2-imperfect-csi}
\begin{aligned}
(\mathcal{P}_{\tt{robust}}'):
\begin{cases}
\min\limits_{\lbrace \pmb{x} \rbrace}
&  \big(\Vert\pmb{x} - \widetilde{\pmb{x}}_{\comm} \Vert + 
	\sigma_{\Delta}'\Vert\pmb{x}\Vert \big) ^2 \\
\textrm{s.t.}
 &  \textcolor{black}{\Vert \pmb{x} \Vert^2 = 1,} \\ 
 & \pmb{x}^H \pmb{F}_n \pmb{x} \leq \frac{\eta}{NL} , \quad \forall n \\
 & \pmb{x} \in \mathcal{B}_{\epsilon}(\pmb{x}_0), \\ 
\end{cases}
\end{aligned}
\end{equation}
and the equivalent real-valued problem is 
\begin{equation}
 \label{eq:problem3-imperfect-csi}
\begin{aligned}
(\bar{\mathcal{P}}_{\tt{robust}}'):
\begin{cases}
\min\limits_{\lbrace \bar{\pmb{x}} \rbrace}&  \big(\Vert\bar{\pmb{x}} - \bar{\widetilde{\pmb{x}}}_{\comm} \Vert + 
	\sigma_{\Delta}'\Vert\bar{\pmb{x}}\Vert \big) ^2  \\
\textrm{s.t.}
 &  \textcolor{black}{\Vert \bar{\pmb{x}} \Vert^2 = 1,} \\ 
 &  \bar{\pmb{x}}^H \bar{\pmb{F}}_n  \bar{\pmb{x}} \leq \frac{\eta}{NL} , \quad \forall n \\
 &  \bar{\pmb{x}} \in \mathcal{B}_{\epsilon}( \bar{\pmb{x}}_0), \\ 
\end{cases}
\end{aligned}
\end{equation}
where $\bar{\widetilde{\pmb{x}}}_{\comm} = \begin{bmatrix}
	\Real({\widetilde{\pmb{x}}}_{\comm})^T & \Imag({\widetilde{\pmb{x}}}_{\comm})^T
\end{bmatrix}^T$ and $\bar{\pmb{x}},\bar{\pmb{x}}_0$ are similarly defined. Following the same approach, we can write the augmented Lagrangian function as  
\begin{equation}
\begin{split}
	&\mathcal{L}_{\rho}^{\tt{robust}}(\bar{\pmb{x}},\pmb{\alpha},\pmb{\beta},\pmb{\gamma}_n,\pmb{u},\pmb{v},\pmb{w}_n) \\
	&=
	\big(\Vert\bar{\pmb{x}} - \bar{\widetilde{\pmb{x}}}_{\comm} \Vert + 
	\sigma_{\Delta}'\Vert\bar{\pmb{x}}\Vert \big) ^2
	+ \pmb{u}^T ( \bar{\pmb{x}} - \pmb{\alpha} ) + \pmb{v}^T ( \bar{\pmb{x}} - \bar{\pmb{x}}_0 - \pmb{\beta}) \\
	&+  \sum\limits_{n=1}^{NL} \pmb{w}_n^T (\bar{\pmb{F}}_n  \bar{\pmb{x}}  - \pmb{\gamma}_n) + \frac{\rho}{2} \Vert \bar{\pmb{x}} - \pmb{\alpha} \Vert^2 + \frac{\rho}{2} \Vert \bar{\pmb{x}} - \bar{\pmb{x}}_0 - \pmb{\beta} \Vert^2 \\
	&+  \frac{\rho}{2} \sum\limits_{n=1}^{NL} \Vert \bar{\pmb{F}}_n  \bar{\pmb{x}}  - \pmb{\gamma}_n \Vert^2.
\end{split}
\end{equation}
Setting the gradient to zero, we can now solve for $\bar{\pmb{x}}^{(m+1)}$ as
\begin{equation}
\begin{split}
\label{eq:gradient-at-x-expanded-imperfect-csi}
	& 2(\Vert \bar{\pmb{x}}^{(m+1)} - \bar{\widetilde{\pmb{x}}}_{\comm} \Vert + \sigma_{\Delta}' \Vert \bar{\pmb{x}}^{(m+1)} \Vert )\\
	& \times (\frac{\bar{\pmb{x}}^{(m+1)} - \bar{\widetilde{\pmb{x}}}_{\comm}}{\Vert \bar{\pmb{x}}^{(m+1)} - \bar{\widetilde{\pmb{x}}}_{\comm} \Vert } + \sigma_{\Delta}' \frac{\bar{\pmb{x}}^{(m+1)}}{\Vert \bar{\pmb{x}}^{(m+1)} \Vert }) + \pmb{u}^{(m)}
	  \\ &  + \pmb{v}^{(m)}  + \rho \sum\limits_{n=1}^{NL} \bar{\pmb{F}}_n^T(\bar{\pmb{F}}_n \bar{\pmb{x}}^{(m+1)} - \pmb{\gamma}_n^{(m)}) + \sum\limits_{n=1}^{NL} \bar{\pmb{F}}_n^T \pmb{w}_n^{(m)}   \\
	&+ \rho (\bar{\pmb{x}}^{(m+1)} - \pmb{\alpha}^{(m)}) 
	+ \rho(  \bar{\pmb{x}}^{(m+1)} - \bar{\pmb{x}}_0 - \pmb{\beta}^{(m)} )  = \pmb{0}.\\
\end{split}
\end{equation}
Notice that a closed form solution in $\bar{\pmb{x}}^{(m+1)}$ is difficult to achieve. Therefore, we resort to a fixed-point iteration type method to solve for $\bar{\pmb{x}}^{(m+1)}$, i.e.
\begin{equation}
\label{eq:fixed-point-iteration}
	\bar{\pmb{x}}^{(m+1)}_{(p+1)}
	=
	\frac{1}{2+3\rho + 2 (\sigma_{\Delta}')^2}
	\phi(\bar{\pmb{x}}^{(m+1)}_{(p)}),
\end{equation}
where 
\begin{equation}
\begin{split}
	\phi(\pmb{x})
	&=
	\Bigg(
	2 \bar{\widetilde{\pmb{x}}}_{\comm} 
	- \pmb{u}^{(m)} 
	- \pmb{v}^{(m)}
	- \sum\limits_{n=1}^{NL}\bar{\pmb{F}}_n \pmb{w}_n^{(m)} + \rho \pmb{\alpha}^{(m)} \\ &
	 + \rho (\bar{\pmb{x}}_0 + \pmb{\beta}^{(m)} ) 
	 + \rho \sum\limits_{n=1}^{NL} \bar{\pmb{F}}_n \pmb{\gamma}_n^{(m)} \\ &
	-2 \sigma_{\Delta}' (\frac{\Vert \pmb{x} \Vert}{\Vert \pmb{x} - \bar{\widetilde{\pmb{x}}}_{\comm} \Vert} (\pmb{x} - \bar{\widetilde{\pmb{x}}}_{\comm} ) + \frac{\Vert \pmb{x} - \bar{\widetilde{\pmb{x}}}_{\comm} \Vert}{\Vert \pmb{x} \Vert} \pmb{x})
	\Bigg).
	\end{split}
\end{equation}
The initialization of the fixed-point iteration in \eqref{eq:fixed-point-iteration} is done as $\bar{\pmb{x}}^{(m+1)}_{(0)} = \pmb{0}$. Note that for the case of perfect \ac{CSI}, i.e. when $\sigma_{\Delta}' = 0$, one iteration of the above fixed-point scheme suffices to converge to the expression in equation \eqref{eq:x-update}. After iterating over $p$, the \ac{ADMM} approach then follows the same steps to update $\pmb{\alpha}^{(m+1)}$ through equation \eqref{eq:alpha-update}, $\pmb{\beta}^{(m+1)}$ via equation \eqref{eq:beta-update},  $\lbrace \pmb{\gamma}_n \rbrace_{n=1}^{NL}$ using equation \eqref{eq:gamma-update}, $\pmb{u}^{(m+1)}$ via equation \eqref{eq:u-update}, $\pmb{v}^{(m+1)}$ by equation \eqref{eq:v-update}, and $\lbrace \pmb{w}_n \rbrace_{n=1}^{NL}$ by equation \eqref{eq:w-update}.

%% file: actions01/LandWaveform.tex
We fix $L = 20$ samples and, unless otherwise stated, we use the orthogonal \ac{LFM} waveform for radar, which can be expressed as 
\begin{equation}
	\pmb{X}_0(m,\ell)
	=
	\frac{1}{\sqrt{NL}}
	\exp(j \frac{2 \pi m}{L} (\ell - 1) )
	\exp(j \frac{ \pi m}{L} (\ell - 1)^2 ).
\end{equation}

%% file: actions01/costisnowaveraged.tex
The obtained cost per iteration shown in Fig. \ref{fig:CostvsIter} is computed as the average of the costs obtained over all the $10^3$ Monte-Carlo trials.

%% file: actions01/sim-cost-convergence.tex
\textcolor{black}{\input{./actions01/sim-cost-convergence-beginning.tex}} 
It can be observed that a lower objective cost is achieved for higher values of $\eta$ at a given constellation size. For example, if QPSK is considered, a cost of about $3$ dB is  when \textcolor{black}{$\eta = 0\dB$} as compared to $\sim -280 $dB when \textcolor{black}{$\eta = 9\dB$} and $\sim -60$ dB when \textcolor{black}{$\eta = 6\dB$}. On the other hand, converging to lower cost values will require additional number of iterations. For example for \textcolor{black}{$\eta = 9\dB$}, the algorithm  iterates for about $550$ iterations to fully converge, as compared to a maximum of $150$ iterations for \textcolor{black}{$\eta = 6\dB$}. 
\textcolor{black}{\input{./actions01/sim-cost-convergence-end.tex}} 

%% file: actions01/sim-cost-convergence-beginning.tex
The simulation demonstrates the convergence behavior per iteration number in terms of MUI energy as a function of $\eta$ and different constellation sizes.

%% file: actions01/sim-cost-convergence-end.tex
For fixed transmit energy, we see that as constellation size increases, the MUI increases as well. For instance, at $\eta=9\dB$, the MUI converges to approximately $-44.2\dB$ for $16-$QAM and to $-32.02\dB$ for $64-$QAM. 

%% file: actions01/sim-papr-ccdf-behaviour-end.tex
Another gap worth highlighting is when the constellation size decreases. For instance, fixing again the  \ac{CCDF} probability to $10^{-2}$, $\eta = 0\dB$ and $\rho = 0.1$, the required $\gamma$ is $3.39\dB$ for QPSK as opposed to $4.19\dB$ for $256-$QAM. This gap decreases for increasing $\eta$ or by decreasing $\rho$. 

%% file: actions01/sim-papr-behaviour.tex
In Fig. \ref{fig:PAPRvsIter}, we plot the $\PAPR$ (in $\dB$) versus the iteration number for different values of $\eta$. Similar to the experiment in Fig. \ref{fig:CostvsIter}, we average the $\PAPR$ of the obtained waveforms over $10^3$ Monte-Carlo trials for each value of $\eta$. Focusing on QPSK, we see that the three curves corresponding to $\eta = 0\dB$, $\eta = 3\dB$ and $\eta = 4.8\dB$  converge to their expected value specified by $\eta$. Moreover, we can observe that the number of iterations required for $\PAPR$ convergence depends on the tuned $\eta$ value. In particular, a lower targeted $\PAPR$ specified by $\eta$ results in more iterations for convergence towards a stable waveform with the desired $\PAPR$. For example, setting $\eta = 4.8 \dB$ necessitates about $30$ iterations to converge to a waveform with a stable $\PAPR$, as opposed to $40$ iterations when the required $\PAPR$ is set to $\eta = 3 \dB$ and $60$ iterations when set to $\eta = 0 \dB$. \input{actions01/sim-papr-behaviour-end.tex}

%% file: actions01/sim-papr-behaviour-end.tex
Interestingly, we see that for a lower constellation size, the \ac{PAPR} tends to settle for a lower value than $\eta$. In particular, when $\eta = 4.8\dB$, the \ac{PAPR} converges to $4.3\dB$ for QPSK, whereas it converges to $4.76\dB$ for $16-$QAM, $64-$QAM and $256-$QAM. As $\eta$ decreases, all constellations converge to the desired \ac{PAPR} value, i.e. $\eta$. 

%% file: actions01/qcqp.tex
The QCQP convex bound is a lower bound in a sense that the constant modulus constraint is relaxed so as the resulting problem is QCQP. See \cite{liu2018toward} for more details.

%% file: actions01/insightRef30_Tradeoff.tex
 It is worth noting that the low \ac{PAPR}-\ac{DFRC} design \cite{hu2022low} does not include a controllable similarity parameter directly in the constraints, but rather as a weighted term in the cost function of \cite{hu2022low}. As a result, we have solved the low \ac{PAPR}-\ac{DFRC} design \cite{hu2022low} for multiple weighting parameters, then chose the weight that corresponds to a desired similarity constraint. We observe that even though the \ac{DFRC}-\ac{PAPR} design in \cite{hu2022low} respects the similarity constraints, it cannot achieve the \ac{AWGN} capacity under the given \ac{PAPR} and similarity constraints.

%% file: actions01/differences-in-pulse-gain-compression.tex
Note that the differences noticed in the pulse gain compression are due to the actual similarity obtained by each of the algorithms. Even though the similarity constraint $\Vert \pmb{x} - \pmb{x}_0 \Vert \leq \epsilon$ is feasible, the \ac{ADMM} method tends to be closer to the constraint's boundary, translating to different pulse gain compressions.

%% file: actions01/SER_imperfect_CSI.tex
In Fig. \ref{fig:SER}, we aim at studying the communication performance of the proposed waveform design, in terms of \ac{SER}, as compared to different algorithms and under the effect of imperfect \ac{CSI}. In this setting, QPSK with $N = 5$ antennas and $K = 2$ communication users are considered. Furthermore, we have set the radar similarity to $\epsilon = 1.25$. In particular, we compare the \ac{SER} of the proposed \ac{ADMM} method compared to \ac{BnB}, SQR-BS, the low \ac{PAPR}-\ac{DFRC} design in \cite{hu2022low} and a benchmark of zero MUI.

At an \ac{SER} level of $10^{-2}$, the \ac{ADMM} waveform design method degrades by only $3.37\dB$ when $\sigma_{\Delta} = -2\dB$ relative to the perfect \ac{CSI} case. Note that \ac{ADMM} is $4.5\dB$ away from the zero MUI benchmark. The gain between the proposed \ac{ADMM} method and all other methods is at least $5\dB$.


\input{actions01/insightRef30_imperfectCSI.tex}

%% file: actions01/insightRef30_imperfectCSI.tex
Interestingly, at an \ac{SER} level of $10^{-1}$, the proposed method degrades with $0.6\dB$ when $\sigma_{\Delta} = -2\dB$, relative to the perfect \ac{CSI} scenario, whereas the low \ac{PAPR}-\ac{DFRC} design \cite{hu2022low} design degrades with $4.9\dB$ and the \ac{BnB} goes beyond a $5\dB$ loss. This again proves the superiority of the \ac{ADMM} waveform design in terms of \ac{SER} and the robustness against channel uncertainties. 

%% file: actions01/acknowledge.tex
The authors would like to thank the anonymous reviewers for their constructive comments, which contributed in improving the manuscript.